\documentclass[showpacs,aps,amsmath,amssymb,nofootinbib,superscriptaddress]{revtex4}
\usepackage{graphicx,floatflt,epsfig,axodraw}
\def\n1{{\cal{N}}_{1i}}

\def\CC5{\left| X^TCX\right|^2}

\def\lpijk{\lambda'_{ijk}}
\def\lpikp{\lambda'_{ikp}}
\def\lpjpk{\lambda'_{jpk}}
\def\lptwtwth{\lambda'_{223}}
\def\lpthtwth{\lambda'_{323}}
\newcommand{\ba}{\begin{array}}    
\newcommand{\ea}{\end{array}}    
\newcommand{\bd}{\begin{displaymath}}    
\newcommand{\ed}{\end{displaymath}}    
\newcommand{\be}{\begin{equation}}    
\newcommand{\ee}{\end{equation}}    
\newcommand{\bea}{\begin{eqnarray}}    
\newcommand{\eea}{\end{eqnarray}}    

\def\issue(#1,#2,#3){{\bf #1}, (#3) #2} 

\def\APP(#1,#2,#3){Acta Phys.\ Polon.\ \issue(#1,#2,#3)}
\def\ARNPS(#1,#2,#3){Ann.\ Rev.\ Nucl.\ Part.\ Sci.\ \issue(#1,#2,#3)}
\def\CPC(#1,#2,#3){Comp.\ Phys.\ Comm.\ \issue(#1,#2,#3)}
\def\CIP(#1,#2,#3){Comput.\ Phys.\ \issue(#1,#2,#3)}
\def\EPJC(#1,#2,#3){Eur.\ Phys.\ J.\ C\ \issue(#1,#2,#3)}
\def\EPJD(#1,#2,#3){Eur.\ Phys.\ J. Direct\ C\ \issue(#1,#2,#3)}
\def\IEEETNS(#1,#2,#3){IEEE Trans.\ Nucl.\ Sci.\ \issue(#1,#2,#3)}
\def\IJMP(#1,#2,#3){Int.\ J.\ Mod.\ Phys. \issue(#1,#2,#3)}
\def\JHEP(#1,#2,#3){J.\ High Energy Physics \issue(#1,#2,#3)}
\def\JPG(#1,#2,#3){J.\ Phys.\ G \issue(#1,#2,#3)}
\def\MPL(#1,#2,#3){Mod.\ Phys.\ Lett.\ \issue(#1,#2,#3)}
\def\NP(#1,#2,#3){Nucl.\ Phys.\ \issue(#1,#2,#3)}
\def\NIM(#1,#2,#3){Nucl.\ Instrum.\ Meth.\ \issue(#1,#2,#3)}
\def\PL(#1,#2,#3){Phys.\ Lett.\ \issue(#1,#2,#3)}
\def\PRD(#1,#2,#3){Phys.\ Rev.\ D \issue(#1,#2,#3)}
\def\PRL(#1,#2,#3){Phys.\ Rev.\ Lett.\ \issue(#1,#2,#3)}
\def\SJNP(#1,#2,#3){Sov.\ J. Nucl.\ Phys.\ \issue(#1,#2,#3)}
\def\ZPC(#1,#2,#3){Zeit.\ Phys.\ C \issue(#1,#2,#3)}

\begin{document} 

\begin{flushright} 
HRI-P08-08-001\\
RECAPP-HRI-2008-009\\
CU-PHYSICS/12-2008
\end{flushright} 

\title{Two-loop neutrino masses with large R-parity violating interactions in
  supersymmetry}

\author{Paramita Dey }
\affiliation{Regional Centre for Accelerator-based Particle Physics,
Harish-Chandra Research Institute, Chhatnag Road, Jhusi, Allahabad
211019, India}
\author{Anirban Kundu}
\affiliation{Department of Physics, University
of Calcutta, 92 A.P.C. Road, Kolkata 700009, India}
\author{Biswarup Mukhopadhyaya}
\affiliation{Regional Centre for Accelerator-based Particle Physics,
Harish-Chandra Research Institute, Chhatnag Road, Jhusi,
Allahabad 211019, India}
\author{Soumitra Nandi}
\affiliation{Regional Centre for Accelerator-based Particle Physics,
Harish-Chandra Research Institute, Chhatnag Road, Jhusi,
Allahabad 211019, India}

\begin{abstract} 
  We attempt to reconcile large trilinear R-parity violating interactions in a
  supersymmetric (SUSY) theory with the observed pattern of neutrino masses
  and mixing. We show that, with a restricted number of such interaction terms
  with the $\lambda'$-type couplings in the range (0.1-1.0), it is possible to
  forbid one-loop contributions to the neutrino mass matrix. This is
  illustrated with the help of a `working example' where an econnomic choice
  of SUSY parameters is made, with three non-vanishing and `large' R-parity
  violating terms in the superpotential. The two-loop contributions in such a
  case can not only generate the masses in the requisite order but can also
  lead us to specific allowed regions of the parameter space.
\end{abstract}



\maketitle



\setcounter{footnote}{0}
\renewcommand{\thefootnote}{\arabic{footnote}}
\section{Introduction}
\label{intro}
Neutrinos are massless to all orders in perturbation theory in the standard
model (SM). However, the ever-accumulating data on solar, atmospheric and
reactor neutrinos challenge us with the inescapable fact that neutrinos are
massive and their physical states are mixtures of the flavour eigenstates
\cite{sno,bando,rean}.  The SM has to be extended for explaining this. The
simplest extension is the inclusion of `sterile' right-handed neutrinos,
whereby neutrinos may either acquire just Dirac masses or, with lepton number
violation, participate in the see-saw mechanism which accounts for their
ultra-light character.

An alternative mechanism is provided by the supersymmetric (SUSY) extension of
the SM with renormalizable $R$-parity ($R_p$) violating terms in the
Lagrangian \cite{SUSY1,godb}.  The fact that baryon and lepton numbers are but
accidentally conserved in the SM entails the possibility of
$R_p=(-1)^{3B+L+2S}$ being violated in SUSY, where $B$, $L$ and $S$ are baryon
number, lepton number and spin respectively. In order to avoid unacceptably
fast proton decay, either $B$ or $L$ must be conserved, while the other may be
violated.  In the latter situation, small Majorana mass terms for neutrinos
(with $\Delta L = 2$) are generated, without the requirement of any additional
fields \cite{godb}. Thus, the neutrino sector may be looked upon as a
motivation for such $L$-violating interactions.

The multiplicative conservation of $R$-parity prevents the lightest SUSY
particle (LSP) from decaying, as $R_p$ equals $+1$ for all SM particles and
$-1$ for the superparticles. All possibilities of $R_p$-violation are
encapsulated in the following terms of the superpotential:
\begin{equation}
W_{\not R} = \lambda_{ijk} L_iL_jE^c_K + \lpijk L_iQ_jD^c_K +
\lambda''_{ijk} U^c_iD^c_jD^c_K + \epsilon_i L_iH_2,
\end{equation}
\noindent
where the first two trilinear terms and the bilinear term are $\Delta L =1$
and the third term is $\Delta B = 1$. Since we are interested in neutrino
masses, let us assume that $B$ is conserved, and that $R_p$ is broken through
$L$-violating couplings only. Moreover, we are neglecting the bilinear terms
$\epsilon_i L_i H_2$ (on which we will comment later), and consider the
trilinear $\lpijk$-type couplings only to illustrate our point. Elaborate
studies in the recent years have led to constraints at various levels on these
couplings \cite{dreiner-1, gautam}. The pertinent gauge-invariant terms
trilinear in particle/sparticle fields are given by
\begin{eqnarray}
\lpijk \left[ {\tilde\nu}^i_L \bar{d}^k_R d^j_L + {\tilde d}^j_L \bar{d}^k_R
  \nu^i_L + ({\tilde d}^k_R)^* (\bar{\nu}^i_L)^c d^j_L - {\tilde e}^i_L
  \bar{d}^k_R u^j_L - {\tilde u}^j_L \bar{d}^k_R e^i_L - ({\tilde d}^k_R)^*
  (\bar{e}^i_L)^c u^j_L \right] + {\rm h.c.}.
\label{lag}
\end{eqnarray}
It is easy to see from above that the $\lpijk$-type couplings (27 of them
altogether) can generate neutrino masses at the loop level, where the
largest contribution comes from $\lambda'_{i33}$. We expect that all of
the entries in the neutrino mass matrix should lie well within 1 eV.
A generic expression for one-loop masses generated in this fashion is 
\cite{oneloop}
\begin{eqnarray}
(m^{1- \rm loop}_\nu)_{ij} \simeq \frac{3}{8\pi^2} m^d_k m^d_p M_{SUSY}
  \frac{1}{{m}^2_{{\tilde q}}} \lpikp\lpjpk,
\label{mass1loop}
\end{eqnarray}
where $m^d_k$ is the down-type quark mass of $k^{\rm th}$ generation,
${m}^2_{{\tilde q}}$ is the (average) squark mass squared, and
$M_{SUSY}~(\sim\mu$, the Higgsino mass parameter) is the effective
scale of SUSY breaking. If the masses thus induced have to answer to
the observed pattern, then a SUSY breaking mass scale of about 500 GeV
would in general imply $\lambda'\sim 10^{-5}-10^{-4}$ \cite{oneloop}.
A similar conclusion follows for $\lambda$-type terms, too.

The question to ask is: are all trilinear R-parity violating couplings thus
destined to be so small, irrespective of all other phenomenological
considerations? For example, will the observation of any process which
requires large values of some $\lambda'$-terms mean that we need some
additional mechanism to explain the neutrino mass pattern? We wish to
demonstrate in this paper that it is not so, so long as one can eliminate the
one-loop contributions but allow two-loop ones, through a limited number of
$\lpijk$-terms. This drastically reduces the number of the $\lambda'$ terms
whose signals may be of interest at the Large Hadron Collider (LHC), be it for
direct observation or through indirect radiative effects. 

Situations where $R_p$-violating two-loop effects can contribute
substantially compared to those at the one-loop level have been
studied in earlier works \cite{borz1}. In contrast, let us assume here
a scenario in which there is a `minimal' set of non-zero {\em large}
($\sim 0.1 - 1.0$) $\lambda'$- type couplings at the weak scale. One
can clearly see from Eq.\ (\ref{mass1loop}) that for such large
$\lambda'$'s, it is impossible to explain the existing neutrino data,
without going into unrealistically high values for $m^2_{{\tilde q}}$,
if both $\lpikp$ and $\lpjpk$ are allowed for the relevant
$\{ij\}$-sets. A way out of this problem would be to postulate this
minimal set of large $\lambda'$'s, of such composition that the above
combinations do not exist, and the relevant interaction terms of Eq.\
(\ref{lag}) contribute to the neutrino mass matrix at the two-loop
level (and beyond) only. At this level, together with the usual loop
suppression factors with respect to the one-loop contributions, there
will be additional suppression coming from the parameters describing
left-right mixing among different flavours in the squark mass matrices
\cite{fcnc} \footnote{Squark mixing parameters can in general occur
along squark propagators, and may enter into one-loop contributions as
well. We will talk about such loops in section \ref{sec2}. In our
analysis however, we have disallowed such combinations of $\lambda'$'s,
and have only retained those which generate neutrino mass terms at the
two-loop level.}. It is thus interesting to see whether these two
suppression factors {\em together} may offset the `largeness' of
$\lambda'$'s, ultimately yielding contributions to neutrino mass matrix
in accordance with the existing neutrino data.

While there are no direct evidences of the Nature favouring any 
particular $R_p$-violating coupling over the others, one may, as a 
starting point, take those that are supported by the low-energy data.
As a case in point, it has recently been advocated in 
\cite{soumitra:bsds} that a minimal set of
three $R_p$-violating couplings can simultaneously explain two
interesting observations in flavour physics. The first one, as shown
by the UTfit Collaboration, is the existence of a sizable deviation of
the $B_s$-$\overline{B}_s$ mixing phase, $\beta_s$, from its SM
expectation, which is close to zero. The second one is the abnormally
large leptonic branching ratio of the $D_s$ meson
\cite{dobrescu-kronfeld}. In ref.\ \cite{soumitra:bsds}, it was found
that one must have large $\lptwtwth$ ($\lpthtwth$) to explain the
recent data on $D_s\to\mu\nu$ ($D_s\to\tau\nu$) \cite{dsdata}, and in
addition either $\lambda'_{212}$ or $\lambda'_{312}$ on similar order,
contributing to the phase in $B_s$ mixing. The $D_s$ anomaly stems
from a very accurate determination of the decay constant, $f_{D_s}$,
on the lattice by the HPQCD Collaboration
\cite{hpqcd}. It has been pointed
out in \cite{lubicz:0807.4605} that further clarifications are needed
on some of the approximations used in \cite{hpqcd}, and prior to that,
it may be advisable to use a more conservative estimate of $f_{D_s}$,
namely, ($250\pm 15$) MeV. Such a value is not in direct conflict with
the experimental number ($273\pm 10$) MeV, and if one wishes to invoke
$R_p$-violation to explain the slight excess, one may use smaller
values of the relevant couplings than those used in
\cite{soumitra:bsds}. On the whole, we take the above result 
as a motivating feature of our analysis, without committing ourselves
too decidedly on any specific numerical values.  

It may be in order to spell out at this stage how general 
our approach is, by re-iterating its main motivation. 
We would like to emphasize that we are not just
attempting to compute two-loop diagrams  contributing to neutrino
masses, which have not been evaluated before. Nor is the sole purpose
of this investigation to account for the claims on $D_s$ decays. The
principal point made by us is that {\em one can reconcile large
R-parity violating couplings and neutrino masses, if only a subset
of all possible couplings of such nature exist.} If there is 
indication of large couplings, the subset must further be determined 
by the impossibility of generating one-loop neutrino masses. Two-loop 
contributions are tenable in such situations, and they can fit the
entire neutrino mass matrix answering to the experimental constraints.
An essential additional ingredient of this mechanism is SUSY flavour
violation through squark mass matrices. We have stressed on identifying
the minimum possible number of $R_p$-and flavour-violating parameters.
This in a way restricts the set of contributing diagrams, but this
feature is {\em characteristic of a minimal choice and not of the
specific couplings chosen, especially if the sfermions of different
flavours are of comparable mass, a feature well-motivated from the
suppression of flavour-changing processes}. Thus this study 
reflects an entire set of possibilities rather than the property of
some specific R-parity violating couplings.
Let us also mention that the values of all $R_p$-violating couplings are 
taken to be those at the electroweak scale and in the mass eigenbasis of
the quarks.

The paper is arranged as follows.  In section \ref{sec2} we discuss
the overall requirements in generating neutrino masses at the two-loop
level only, using $R_p$ violating couplings of the
$\lambda'$-type. Some features of the two-loop contributions are
outlined in section \ref{sec3}, while in section \ref{sec4} we test
the validity of the scheme on a numerical basis. Section \ref{sec5} is
on some correlated signals of the relevant couplings that may be
tested at the LHC. We summarize and conclude in section
\ref{sec6}. Some representative expressions related to the loop
integrals are included in the Appendix.

\section{The parameters relevant for two-loop effects}
\label{sec2} 
Let us try to identify a minimal set of parameters that are required
to generate a neutrino mass matrix at {\em no less than the two-loop
level}. Of course, one requires a set of non-zero $\lambda'$ which can
be allowed to lie in the range $0.1 - 1$. As will be explained below,
one further requires the parameters controlling flavour violation in
the squark sector in order to generate the mass matrix in a way
consistent with observations.

Next, we recall the pattern of the three-family neutrino mass matrix in the
flavour basis, assuming, without any loss of generality, that the charged
lepton mass matrix is diagonal in this basis. 
The constraints on the mixing angles are \cite{pdg} 
\begin{equation}
\sin^2(2\theta_{12}) = 0.86^{+0.03}_{-0.04} \ \Rightarrow \ \theta_{12} 
= (33.89\pm 1.44)^\circ\,,\ \ \ \sin^2(2\theta_{23}) > 0.92 \ 
\Rightarrow \ \theta_{23} > 36.8^\circ\,,
\end{equation}
and $\sin^2(2\theta_{12}) < 0.19$. We assume the bilarge mixing scheme so that 
$\theta_{23}=\pi/4$ and $\theta_{13}=0$  \cite{massmatrix},
\begin{eqnarray}
M_{\nu} =\ \left(\! \begin{array}{ccc}
m_1c^2+m_2s^2 & \frac{cs}{\sqrt 2}(-m_1+m_2) & \frac{cs}{\sqrt 2}(m_1-m_2) \\
\frac{cs}{\sqrt 2}(-m_1+m_2) & \frac{1}{2}(m_1s^2+m_2c^2+m_3) &
\frac{1}{2}(-m_1s^2-m_2c^2+m_3) \\
\frac{cs}{\sqrt 2}(m_1-m_2) &
\frac{1}{2}(-m_1s^2-m_2c^2+m_3) & \frac{1}{2}(m_1s^2+m_2c^2+m_3) 
\end{array}\!\right)\,,
\label{matrix}
\end{eqnarray}
where $m_1$, $m_2$, $m_3$ are the mass eigenvalues, and
$s=\sin\theta_{12}$, $c=\cos\theta_{12}$, $\theta_{ij}$ being the
mixing angle between the $i$th and the $j$th family. From this matrix
one can easily take up the specific scenarios of normal ($m_3 >> m_2
\gtrsim m_1$) or inverted ($m_2 \gtrsim m_1 >> m_3$) hierarchy or that
of degenerate neutrinos ($m_1 \simeq m_2 \simeq m_3$). One can take
$m_1 = 0$ for normal hierarchy (NH) and $m_3 = 0$ for inverted
hierarchy (IH), without any loss of generality.  In the case of NH,
the existing data require (at 95\% confidence limit)
\begin{eqnarray} 
m^2_2 = (7.60 \pm 0.35) \times 10^{-5}~{\rm eV}^2, ~ \left| m^2_3-m^2_2\right|
= (2.50 \pm 0.27) \times 10^{-3}~{\rm eV}^2,
\label{cons1}
\end{eqnarray}
and $s^2=0.3$. The corresponding numbers for IH and
degenerate neutrinos (DN) are
\begin{eqnarray} 
m^2_2 - m^2_1 = (7.60 \pm 0.35) \times 10^{-5} && \!\!\!\! {\rm eV}^2,
~
\left| m^2_2-m^2_3\right| \simeq \left| m^2_1 - m^2_3 \right| = (2.50 \pm 0.27)
\times 10^{-3}~{\rm eV}^2, \nonumber \\
\end{eqnarray}
and
\begin{eqnarray}
&{}& m_1 \simeq m_2 \simeq m_3 \simeq O(10^{-1}) ~{\rm eV}
\label{cons2}
\end{eqnarray}
respectively \cite{bando2}. 

Let us first try to understand intuitively the 
properties of the `minimal set'. 
\begin{itemize}
\item
There must be no less than three $\lambda'$ type couplings,
each with a different leptonic index, for the three neutrinos. 
\item
To prevent mass
generation at one-loop, couplings like $\lambda'_{ijj}$, which generate
diagonal entries of the neutrino mass matrix,  are forbidden. 
\item
Similarly, 
combinations like $\lambda'_{ikl}\lambda'_{jlk}$ are forbidden to prevent
the off-diagonal entries at the one-loop level. 
\item
In fact, $\lambda'_{ikl}
\lambda'_{jmk}$ combinations are also not allowed, since they can generate
one-loop masses with the mass insertion $\delta_{ml}^{LR}$. 
\end{itemize}
This leaves us with a limited number of possible choices.

As already discussed in Section \ref{intro}, our choice of the supersymmetric
scenario is partially motivated by the explanation of the results on $D_s$
decays \footnote{This is a partial motivation because, as we will show later,
  the allowed values of $\lambda'_{223}$ and $\lambda'_{323}$ result only in a
  marginal enhancement of the $D_s$ leptonic branching ratio. However, it is
  better to be cautious about the HPQCD lattice result.}.  We thus include
$\lptwtwth$ and $\lpthtwth$ in our minimal set of $\lambda'$-type couplings,
and propose that their values be allowed to be large, consistent with the
individual constraints.  It is easy to see from the relevant interaction terms
(the second and third terms of Eq.\ (\ref{lag})) that we need one more
$\lpijk$ with $i=1$, in order to have contributions to the elements in the
first row and the first column of $M_\nu$.  The choices that we thus have are
$\lambda'_{112}$, $\lambda'_{121}$, $\lambda'_{113}$, $\lambda'_{131}$ and
$\lambda'_{123}$.
\begin{figure}
\begin{center}
\begin{picture} (300,56)(0,0)
\Vertex(180,10){1.5}
\Vertex(120,10){1.5}
\Text(150,40)[]{$\otimes$}
\ArrowLine(80,10)(120,10)
\ArrowLine(120,10)(180,10)
\ArrowLine(180,10)(220,10)
\DashArrowArcn(150,10)(30,90,0){4}
\DashArrowArc(150,10)(30,90,180){4}
\Text(110,30)[]{$\tilde{d}_p$}
\Text(190,30)[]{${\tilde{d}}^*_{p'}$}
\Text(100,0)[]{$\nu_i$}
\Text(150,0)[]{$d_k$}
\Text(200,0)[]{$\nu^c_j$}
\end{picture}
\caption{ A typical one-loop diagram contributing to neutrino masses. It should
be remembered that corresponding to each such diagram there is one with $\nu_i$
and $\nu^{c}_j$ flipped. }
\end{center}
\end{figure}
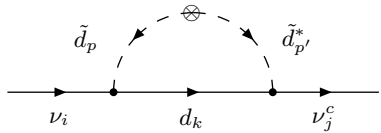

Let us clarify the last criterion mentioned above.
As a first choice, let us choose $\lambda'_{112}$. It is then easy to see from
Figure 1 that there are non-vanishing one-loop contributions to the $(1,2)$
and $(1,3)$ elements of the neutrino mass matrix. This is because of the fact
that the quark and squark mass matrices of the same charge are not in general
diagonal simultaneously; the evolution of the squark mass parameters from the
high scale of SUSY breaking always tend to destroy such alignment. The
resulting possibility of a flavour transition as well as a chirality flip
along the down-type squark propagator allows one to obtain some one-loop
contributions to the neutrino mass matrix. These contributions are driven by a
parameter $\delta^{LR}_{13}$ of mass-squared dimension, which is basically the
corresponding off-diagonal term in the down-type squark mass matrix
\footnote{Our convention is different from, say, that of \cite{fcnc}.  While
  our $\delta$, which is of mass-squared dimension, is identical to their
  $\Delta$, the $\Delta$ parameters that we subsequently introduce are based
  on a different scaling. We have checked that the existing numerical
  constraints are all satisfied.}.  Such diagrams are not suppressed enough to
balance the large values ($O(0.1)$) of $\lptwtwth$ and $\lpthtwth$ and give
admissibly small entries for the $(1,2)$ and $(1,3)$ elements of the neutrino
mass matrix.  So, a non-zero $\lambda'_{112}$ will not normally serve our
purpose. Besides, the phenomenological constraint on $\lambda'_{112}$ makes it
an inappropriate candidate for the demonstration of the effects of large $R_p$
violating interactions.  Following similar arguments, the choice of
$\lambda'_{131}$ should be abandoned in our minimal set of $\lambda'$'s.

On the contrary, none among $\lambda'_{121}$, $\lambda'_{113}$ and
$\lambda'_{123}$ can give one-loop contributions to the neutrino mass matrix.
So in principle any one of them can be included in the minimal set together
with $\lptwtwth$ and $\lpthtwth$, to generate neutrino mass at two-loop level.
Note that the choice of $\lambda'_{123}$ puts a single squark mixing parameter
at our disposal, namely, the one describing the second-and third-family squark
mixing ($\delta_{23}^{LR}$).  Thus we would have a set of four independent
parameters: $\lptwtwth$, $\lpthtwth$, $\lambda'_{123}$ and $\delta_{23}^{LR}$.
However, as will be evident from our numerical results in Section \ref{sec4},
it is difficult to fit the six independent elements of $M_{\nu}$ with
experimental data with just these four parameters.

Choosing $\lambda'_{113}$, on the other hand, will involve $\delta_{13}^{LR}$,
the first- and third-family squark mixing parameter, for generating the
elements of $M_{\nu}$ in the first row and first column, and $\delta_{23}$ for
the rest of the matrix elements (from now on, we will drop the chirality
superscript on the $\delta$s, since the only type that we will ever be
interested in are those of the $LR$ type in the down-squark sector).  This
means that for this choice we have a set of five independent parameters
comprising of ($\lptwtwth$, $\lpthtwth$, $\lambda'_{113}$, $\delta_{23}$ and
$\delta_{13}$). In Section \ref{sec4} we will see that in this case we are
able to fit elements of $M_{\nu}$ with the existing constraints. In a similar
way, the choice of $\lambda'_{121}$ also leads to the same number of
independent parameters. However, for the latter choice, some two-loop
contributions would be suppressed further by the ratio $m_s/m_b$, making the
two-loop effects undesirably small, as we shall see in Section \ref{sec4}.

Thus, our selected parameter space consists of a minimal set of three $O(0.1)$
$\lambda'$'s, namely $\lptwtwth$, $\lpthtwth$ and $\lambda'_{113}$, and two
non-zero squark mixing parameters $\delta_{13}$ and $\delta_{23}$, generating
neutrino masses at the two-loop level. All other parameters are set to be zero
at the weak scale.  Also, we will work under the assumption of all the
$\lambda'$'s being real.

In our calculation, we scale the squark mixing parameter $\delta_{ij}$ by the
factor $m_b M_{SUSY}$, and define a dimensionless parameter ${\Delta_{ij}} =
\delta_{ij}/(m_b M_{SUSY})$, with the already specified connotation
$\Delta_{ij} = \Delta^{LR}_{ij}$. The various loop contributions which involve
flavour violation and require a chirality flip in the (down-type) squark
propagator are expressed in terms of $\Delta_{13}$ and $\Delta_{23}$.

The coupling $\lambda'_{113}$ is bounded from charged current universality
\cite{barger-giudice-han} as well as processes like $\pi^+\to e^+ \nu_e$. Here
we use a 99\% confidence level bound of $|\lambda'_{113}|\leq 0.15$. As we
shall see, this relative smallness of $\lambda'_{113}$ leads to a distinct
preference of the NH scenario of neutrino masses over IH or DN.

$\lambda'_{223}$ and $\lambda'_{323}$, the other two couplings, can be large,
even $O(1)$. We have checked that the recent CLEO constraint on lepton flavour
violation in $\Upsilon\to \mu\tau$ \cite{cleo:lfv} is consistent with this
upper limit.

Ref.\ 2 of \cite{dreiner-1} quotes a weak scale bounds of $|{\lambda'}_{113}|,
|{\lambda'}_{i23}|\leq 0.39$. These limits arise from the need to prevent
tachyonic sneutrinos even at the Grand Unified Theory (GUT) scale
\cite{decarlos}. The maximum value at the GUT scale is driven by the input
parameters; for the set known as SPS1a, this comes out to be about 0.13. When
run down at the $M_Z$ scale, the coupling increases threefold and the bound
becomes $0.39$.  One can easily relax this bound for other choices of the GUT
scale input parameters.

The dimensionless parameters $\Delta_{ij}$ can be constrained from various
flavour-changing neutral current (FCNC) processes. For those that we are
interested in, $\Delta_{13}$ is constrained from $B^0$-$\overline{B}{}^0$
mixing to be less than 5.2, and $\Delta_{23}$ is constrained from the
inclusive $b\to s \gamma$ branching ratio to be less than 1.0.

Let us mention again that this is just one of several possible choices. 
Following the rules laid down earlier, one must have three $\lambda'$
type couplings and two $\delta$-type squark mixing parameters. However, some
of the possible choices are extremely constrained from data. For example, the
choice of $\lambda'_{121}$, $\lambda'_{221}$, $\lambda'_{323}$, $\delta_{21}^{LR}$
and $\delta_{23}^{LR}$ is severely restricted by the absence of leptonic
flavour-violating decays $\pi^0\to e\mu$, $\phi\to e\mu$, $B\to e(\mu)\tau$
etc.

\section{The two-loop contributions}
\label{sec3}
Having shown that there are no one-loop contributions to the neutrino mass
matrix $M_\nu$, let us enlist and compute the two-loop contributions that are
driven by the three nonzero $\lambda'$ type couplings and two $\delta$
parameters. We work in the 't Hooft-Feynman gauge.
\begin{figure}
\begin{center}
\begin{picture} (300,250)(0,-200)
\Vertex(180,10){1.5}
\Vertex(120,10){1.5}
\Vertex(80,10){1.5}
\Vertex(220,10){1.5}
\Text(150,80)[]{$\otimes$}
\ArrowLine(40,10)(80,10)
\ArrowLine(80,10)(120,10)
\ArrowLine(120,10)(180,10)
\ArrowLine(180,10)(220,10)
\ArrowLine(220,10)(260,10)
\DashArrowArcn(150,10)(70,90,0){8}
\DashArrowArc(150,10)(70,90,180){8}
\DashCArc(150,10)(30,180,0){4}
\Text(90,70)[]{$\tilde{d}_p$}
\Text(210,70)[]{${\tilde{d}}^*_{p'}$}
\Text(60,0)[]{$\nu_i$}
\Text(100,0)[]{$d_k$}
\Text(150,0)[]{$u_n$}
\Text(200,0)[]{$d_{k'}$}
\Text(240,0)[]{$\nu^c_j$}
\Text(150,-30)[]{$W^+, \phi^+, H^+$}
\Text(150,-55)[]{(a)}
\Vertex(40,-100){1.5}
\Vertex(0,-100){1.5}
\Vertex(-40,-100){1.5}
\Vertex(80,-100){1.5}
\Text(0,-60)[]{$\otimes$}
\ArrowLine(-80,-100)(-40,-100)
\ArrowLine(-40,-100)(0,-100)
\ArrowLine(0,-100)(40,-100)
\ArrowLine(40,-100)(80,-100)
\ArrowLine(80,-100)(120,-100)
\DashArrowArcn(0,-100)(40,90,0){4}
\DashArrowArc(0,-100)(40,90,180){4}
\DashCArc(40,-100)(40,180,0){4}
\Text(-40,-60)[]{$\tilde{d}_p$}
\Text(40,-60)[]{${\tilde{d}}^*_{p'}$}
\Text(-60,-110)[]{$\nu_i$}
\Text(-20,-110)[]{$d_k$}
\Text(20,-110)[]{$u_{k'}$}
\Text(60,-110)[]{$\ell^c_j$}
\Text(100,-110)[]{$\nu^c_j$}
\Text(40,-150)[]{$\phi^+, H^+$}
\Text(30,-175)[]{(b)}
\Vertex(300,-100){1.5}
\Vertex(260,-100){1.5}
\Vertex(220,-100){1.5}
\Vertex(340,-100){1.5}
\ArrowLine(180,-100)(220,-100) \ArrowLine(220,-100)(260,-100)
\ArrowLine(260,-100)(300,-100) \ArrowLine(300,-100)(340,-100)
\ArrowLine(340,-100)(380,-100) \DashArrowArcn(300,-100)(40,90,0){4}
\DashArrowArc(300,-100)(40,90,180){4} \DashCArc(260,-100)(40,180,0){4}
\Text(260,-60)[]{$\tilde{d}_k$} \Text(340,-60)[]{${\tilde{d}}^*_{p'}$}
\Text(200,-110)[]{$\nu_i$} \Text(240,-110)[]{$\ell_i$}
\Text(280,-110)[]{$u_{p}$} \Text(320,-110)[]{$d_{k'}$}
\Text(360,-110)[]{$\nu^c_j$} \Text(260,-150)[]{$W^+, \phi^+, H^+$}
\Text(290,-175)[]{(c)} 
\end{picture} 
\caption{Two-loop diagrams that
  make leading contributions to neutrino masses. The contributions in the
  three different diagrams are proportional to (a)
  $\lambda'_{ipk}\lambda'_{jk'p'}\Delta^{LR}_{pp'}V_{u_n d_k}V^{\ast}_{u_n
    d_{k'} }$ (b) $\lambda'_{ipk}\lambda'_{jk'p'}\Delta^{LR}_{pp'}V_{u_{k'}
    d_k}$ (c) $\lambda'_{ipk}\lambda'_{jk'p'}\Delta^{LR}_{kp'}V_{u_p d_{k'}}$.
  For our choice of $\lambda'$'s in (c), $k = p' = 3$ and thus $\Delta_{kp'} =
  1 $, so that there is no squark flavour violation. The flipped diagrams,
  too, will contribute as usual.}
\end{center}
\end{figure}
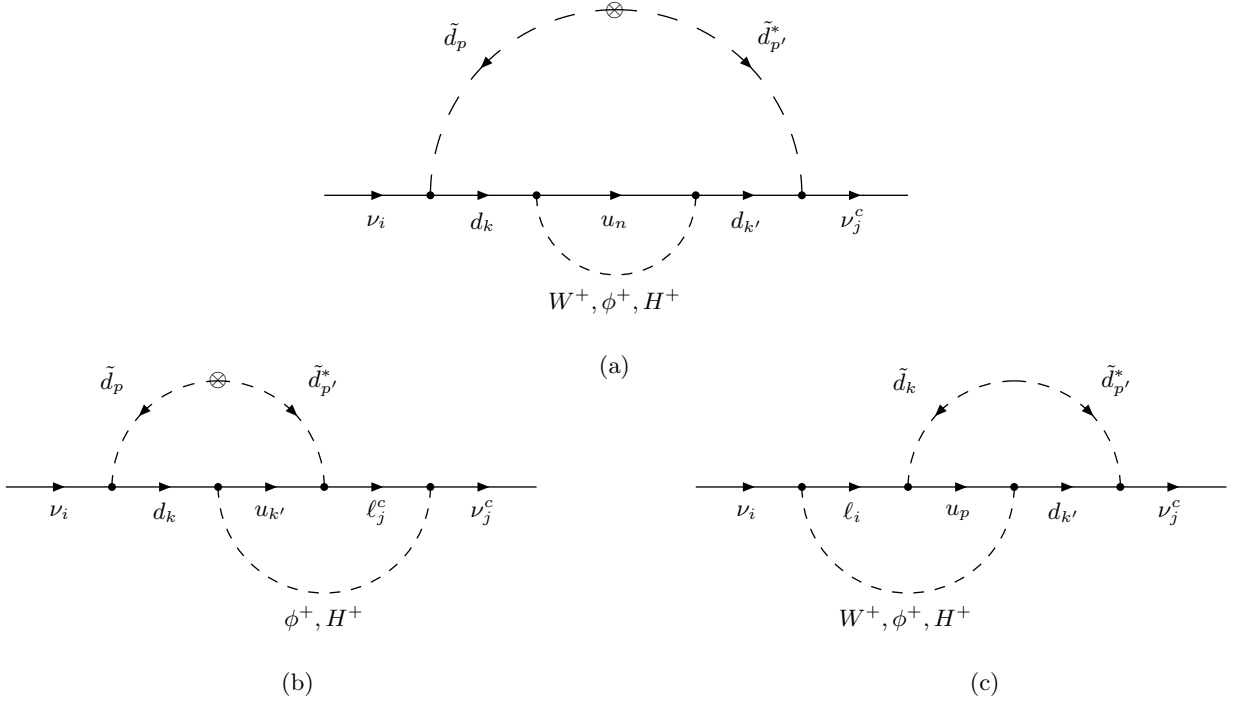 

Although the individual diagrams are divergent, the very fact that
there is no counterterm at the tree-level for the interactions generated at 
higher loop levels
immediately tells us that the end result is finite. This
is ensured when all diagrams including all possible fields and their
superpartners are taken into account.

Figure 2 represents three classes of diagrams that turn out to be dominant in
our study. A full list of generic expressions for the loop-factors arising
from these diagrams is provided in the Appendix. The amplitude corresponding
to diagram 2(a) for the $(1,1)$ and $(i,j)$ elements of $M_{\nu}$, where
$i,j=2,3$, is found to be
\begin{eqnarray}
 \left[M^{\rm 2a}_{\nu}\right]_{11} &\sim&  \frac{m_dm_b}{(m^2_d-m^2_b)}
{\Delta}_{13} \xi_t, \nonumber \\
 \left[M^{\rm 2a}_{\nu}\right]_{ij} &\sim& \frac{m_sm_b}{(m^2_s-m^2_b)} 
{\Delta}_{23} \xi'_t
\label{golgol}
\end{eqnarray}
respectively, where $\xi_t = V^*_{ts}V_{tb}$, and $\xi'_t = V^*_{td}V_{tb}$.
The loop functions have been left out of these expressions.  For the $(1,2)$
and $(1,3)$ elements of $M_{\nu}$, the contribution from diagram 2(a) contains
two separate parts proportional to the two factors written above, along with
the appropriate loop functions multiplying each of them.

The amplitude for diagram 2(b) vanishes when mediated by $W^{\pm}$ (but not
the charged Higgs or Goldstone field), which follows from the details of
$\gamma$-matrix algebra.  For the diagonal entries of the neutrino mass
matrix, this diagram yields
\begin{eqnarray}
\left[M^{\rm 2b}_{\nu}\right]_{11} &\sim& V_{ub} \frac{m_bm_d}{M^2_W} 
{\Delta}_{13} x_e x_u, \nonumber\\ 
 \left[M^{\rm 2b}_{\nu}\right]_{22} &\sim&  V_{cb} \frac{m_bm_s}{M^2_W} 
{\Delta}_{23} x_\mu x_c, \nonumber\\ 
  \left[M^{\rm 2b}_{\nu}\right]_{33} &\sim&  V_{cb} \frac{m_bm_s}{M^2_W}
{\Delta}_{23} x_\tau x_c
\label{ktfv}
\end{eqnarray}
where $x_a=m^2_a/M^2_W$ for $a=e,\mu, \tau, u, c$. Each of the off-diagonal
matrix elements (1,2), (1,3) and (2,3) is a sum of two terms which are
respectively proportional to the first and second, first and third and second
and third factors written above, again with the corresponding loop functions.
In general, being proportional to the squares of lepton masses, the
contribution of diagram 2(b) are suppressed compared to those of Figure 2(a).
{\em Diagram 2(c) is particularly interesting, since there is no flavour
  change required along the internal squark-line in this diagram. Thus, these
  diagrams do not have the suppression by $\Delta$-factors.} Nevertheless,
this diagram has an overall lepton mass dependence. So ultimately it
contributes more than diagram 2(b), but less than 2(a). For the diagonal
entries of the neutrino mass matrix from 2(c),
\begin{eqnarray}
 \left[M^{\rm 2c}_{\nu}\right]_{11} &\sim&  V^*_{ud} x_u m_e, \nonumber\\ 
 \left[M^{\rm 2c}_{\nu}\right]_{22} &\sim&  V^*_{cs} x_c m_\mu, \nonumber\\ 
 \left[M^{\rm 2c}_{\nu}\right]_{33} &\sim&  V^*_{cs} x_c m_\tau 
\label{noktfv}
\end{eqnarray}
while, just as before, each of the off-diagonal entries $M_\nu(i,j)$
separately contains two terms which are proportional to the $i$-th and $j$-th
factors respectively of Eq.\ (\ref{noktfv}). It is thus clear from equations
(\ref{golgol}), (\ref{ktfv}), and (\ref{noktfv}), that the contribution from
diagrams of type 2(a) dominate over the others for the elements in the first
row and first column of $M_{\nu}$, while for the other elements, these are
more or less of the same order.

We have worked with such a choice of the electroweak symmetry breaking sector
that $\tan\beta$ = 10 (where $\tan\beta$ is the ratio of the two Higgs vacuum
expectation values) and the charged Higgs mass is 500 GeV.  The charged Higgs
contributions are found to be suppressed with respect to the ones discussed
above. In a similar manner, the loops involving charginos and neutralinos are
found to be of subleading nature, as their presence would imply additional
squark and slepton propagators, leading to bigger suppression factors under
our choice of mass ($\simeq$500 GeV) for all squarks and sleptons. It is
therefore legitimate to illustrate our main points leaving out such diagrams.
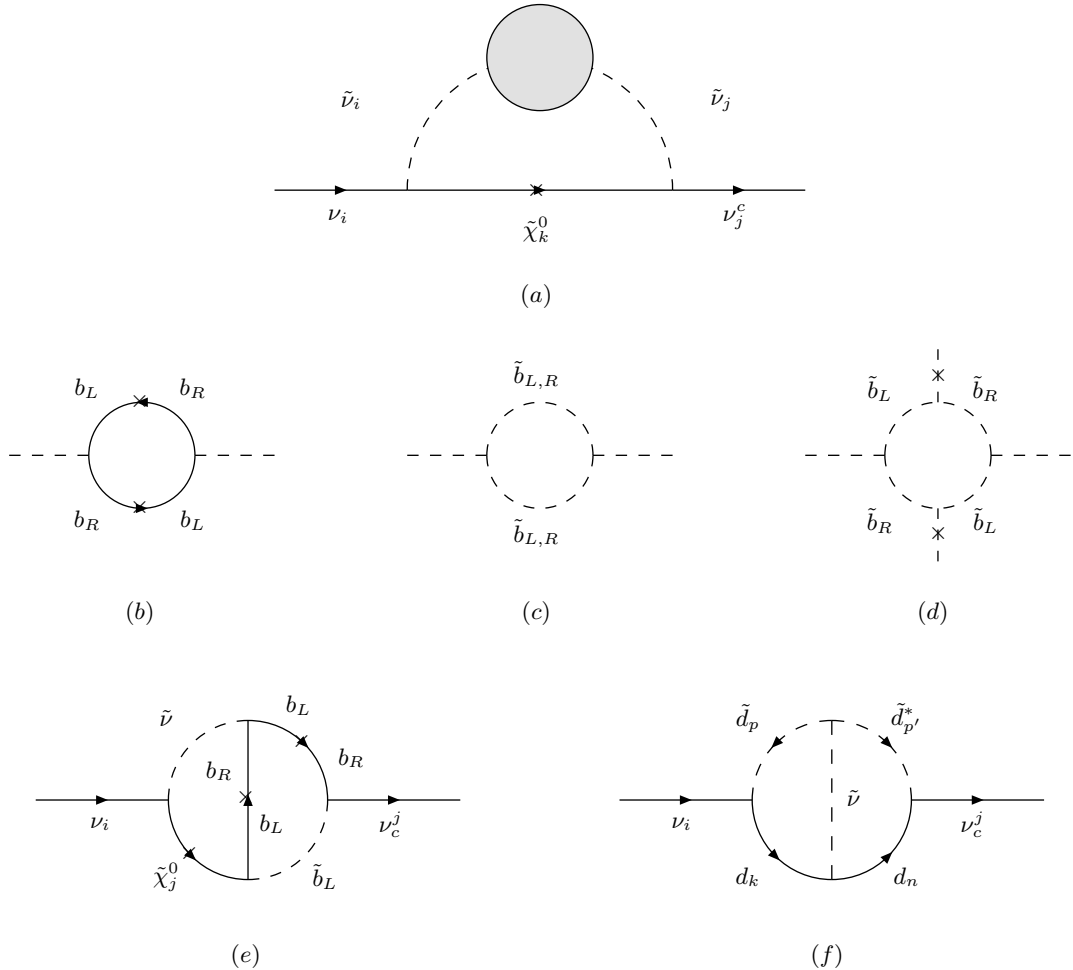
\begin{figure}
\begin{center}
\begin{picture} (600,370)(100,-350)
\ArrowLine(250,-20)(300,-20)
\ArrowLine(300,-20)(400,-20)
\ArrowLine(400,-20)(450,-20)
\DashArrowArc(350,-20)(50,0,180){4}
\GOval(350,30)(20,20)(0){0.882}
\Text(275,-30)[]{{{$\nu_i$}}}
\Text(350,-35)[]{{{${\tilde{\chi}}^0_k$}}}
\Text(425,-30)[]{{{$\nu^c_j$}}}
\Text(280,14)[]{{{${\tilde{\nu}}_i$}}}
\Text(420,14)[]{{{${\tilde{\nu}}_j$}}}
\Text(350,-20)[]{{{$\times$}}}
\Text(350,-60)[]{{{$(a)$}}}
\DashLine(150,-120)(180,-120){4}
\ArrowArc(200,-120)(20,0,180)
\ArrowArc(200,-120)(20,180,360)
\DashLine(220,-120)(250,-120){4}
\Text(200,-100)[]{$\times$}
\Text(200,-140)[]{$\times$}
\Text(180,-95)[]{$b_L$}
\Text(180,-145)[]{$b_R$}
\Text(220,-95)[]{$b_R$}
\Text(220,-145)[]{$b_L$}
\Text(200,-180)[]{{{$(b)$}}}
\DashLine(300,-120)(330,-120){4}
\DashCArc(350,-120)(20,0,180){4}
\DashCArc(350,-120)(20,180,360){4}
\DashLine(370,-120)(400,-120){4}
\Text(350,-90)[]{$\tilde{b}_{L,R}$}
\Text(350,-150)[]{$\tilde{b}_{L,R}$}
\Text(350,-180)[]{{{$(c)$}}}
\DashLine(450,-120)(480,-120){4}
\DashCArc(500,-120)(20,0,180){4}
\DashCArc(500,-120)(20,180,360){4}
\DashLine(520,-120)(550,-120){4}
\DashLine(500,-100)(500,-80){4}
\DashLine(500,-140)(500,-160){4}
\Text(502,-90)[]{$\times$}
\Text(502,-150)[]{$\times$}
\Text(480,-95)[]{$\tilde{b}_L$}
\Text(480,-145)[]{$\tilde{b}_R$}
\Text(520,-95)[]{$\tilde{b}_R$}
\Text(520,-145)[]{$\tilde{b}_L$}
\Text(500,-180)[]{{{$(d)$}}}
\ArrowLine(160,-250)(210,-250)
\ArrowLine(270,-250)(320,-250)
\DashCArc(240,-250)(30,90,180){4}
\DashCArc(240,-250)(30,270,360){4}
\ArrowArc(240,-250)(30,180,270)
\ArrowArcn(240,-250)(30,90,0)
\ArrowLine(240,-280)(240,-220)
\Text(240,-250)[]{$\times$}
\Text(218.79,-271.21)[]{$\times$}
\Text(261.21,-228.79)[]{$\times$}
\Text(185,-260)[]{$\nu_i$}
\Text(295,-260)[]{$\nu^j_c$}
\Text(210,-220)[]{$\tilde{\nu}$}
\Text(210,-280)[]{$\tilde{\chi}^0_j$}
\Text(260,-215)[]{$b_L$}
\Text(280,-235)[]{$b_R$}
\Text(230,-240)[]{$b_R$}
\Text(250,-260)[]{$b_L$}
\Text(270,-280)[]{$\tilde{b}_L$}
\Text(240,-310)[]{{{$(e)$}}}
\ArrowLine(380,-250)(430,-250)
\DashArrowArc(460,-250)(30,90,180){4}
\DashArrowArcn(460,-250)(30,90,0){4}
\ArrowLine(490,-250)(540,-250) 
\ArrowArc(460,-250)(30,180,270)
\ArrowArc(460,-250)(30,270,360)   
\DashLine(460,-280)(460,-220){6}
\Text(405,-260)[]{$\nu_i$}
\Text(515,-260)[]{$\nu^j_c$}
\Text(430,-220)[]{$\tilde{d}_p$}
\Text(430,-280)[]{$d_k$}
\Text(490,-280)[]{$d_n$}
\Text(470,-250)[]{$\tilde{\nu}$}
\Text(490,-220)[]{$\tilde{d}^*_{p'}$}
\Text(460,-310)[]{{{$(f)$}}}
\end{picture}
\caption{Additional two-loop diagram that will not contribute in our
case. $(b)$, $(c)$, $(d)$, correspond to the blob shown in
$(a)$. Contributions from $(f)$ require trilinear $L$-violating soft
terms in the scalar potential.}
\end{center}
\end{figure}
Additional diagrams have been taken into account in earlier works dealing with
two-loop neutrino masses in R-parity violating SUSY \cite{borz1}.
Representative diagrams of this type are shown in Figure 3. 
The reasons for not taking these contributions into account, without
losing generality in our approach, are as follows:

\begin{itemize}
\item The contributions from diagrams 3(a) depend on the splitting
between CP-even and CP-odd sneutrino states. That requires added
theoretical inputs which are not present in our study.
\item Even when one goes beyond the minimal set of R-parity violating
interactions, diagrams of the kind shown in 3(e) cannot contribute
without there being contributions at one-loop, whose absence is
precisely the theme of our work.
\item Diagrams 3(f) require additional assumptions about
soft trilinear terms with $\Delta L = 1$ in the scalar potential.
\end{itemize}

\section{Results and discussion}
\label{sec4}

We have five parameters, namely, $\lambda'_{113}$, $\lambda'_{223}$,
$\lambda'_{323}$, $\Delta_{23}$ and $\Delta_{13}$, with which to fit the
neutrino mass matrix $M_\nu$ to generate the required mass hierarchies. 
Here, as we have already defined, $\Delta_{ij}=\delta_{ij}/m_b M_{SUSY}$.
These parameters, along with 
\begin{eqnarray}
&& m_t =172.5 ~{\rm GeV}\,, \ \ \ m_b = 4.5 ~{\rm GeV}\,, \ \ \ 
|V_{td}|=(8.12 \pm 0.88) \times 10^{-3}\,, \ \ \ 
|V_{ts}|=(40.67 \pm 1.30) \times 10^{-3}\,, \nonumber \\
&& |V_{cb}|=(40.8 \pm 0.6) \times 10^{-3}\,,\ \ \  
\sin 2\beta_d=0.755 \pm 0.040\,, \ \ \ \theta_{12} = (33.89\pm 1.44)^\circ\,,
\label{mnuconstr}
\end{eqnarray}
\noindent
(where $\beta_d = \arg(V_{td}^\ast)$), all sparticle masses (including 
$H^\pm$, and all sleptons and squarks, and hence $M_{SUSY}$) 
at 500 GeV, and $\tan\beta=10$,
essentially determine the entries of $M_\nu$ as shown in the Appendix.  

We vary the SM inputs over their allowed ranges, and the five parameters
$\lambda'$ and $\Delta$ over the range 0.0-1.0, to see whether any simultaneous
solution to the $M_\nu$ constraints exist. We take all the $\lambda'$s and
$\Delta$s to be real
and positive. In fact, there are four independent parameters, and not five,
that need to be varied.  
The reason lies in the neutrino mass matrix $M_\nu$,
whose (2,2) are (3,3) elements are equal for $\theta_{13}=0$ and differ very
slightly for small $\theta_{13}$. The relevant amplitudes, being completely
identical in the leptonic part, imply $\lambda'_{223}\approx \lambda'_{323}$.
Thus, essentially, we have four free parameters, namely, $\lambda'_{113}$,
$\lambda'_{223}$, $\Delta_{13}$ and $\Delta_{23}$.

\begin{figure}[htbp]
\parbox{8cm}{\epsfig{file=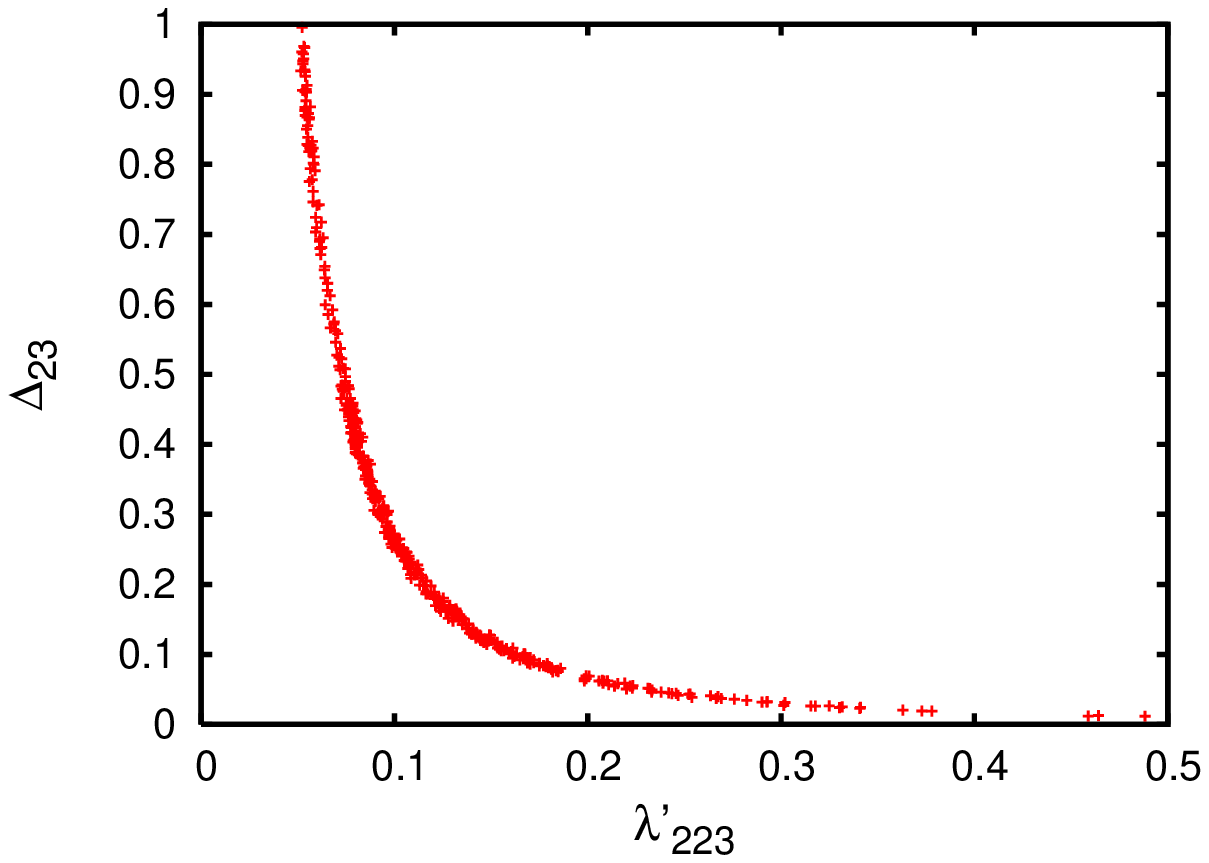,width=8cm,angle=0}}
\parbox{8cm}{\epsfig{file=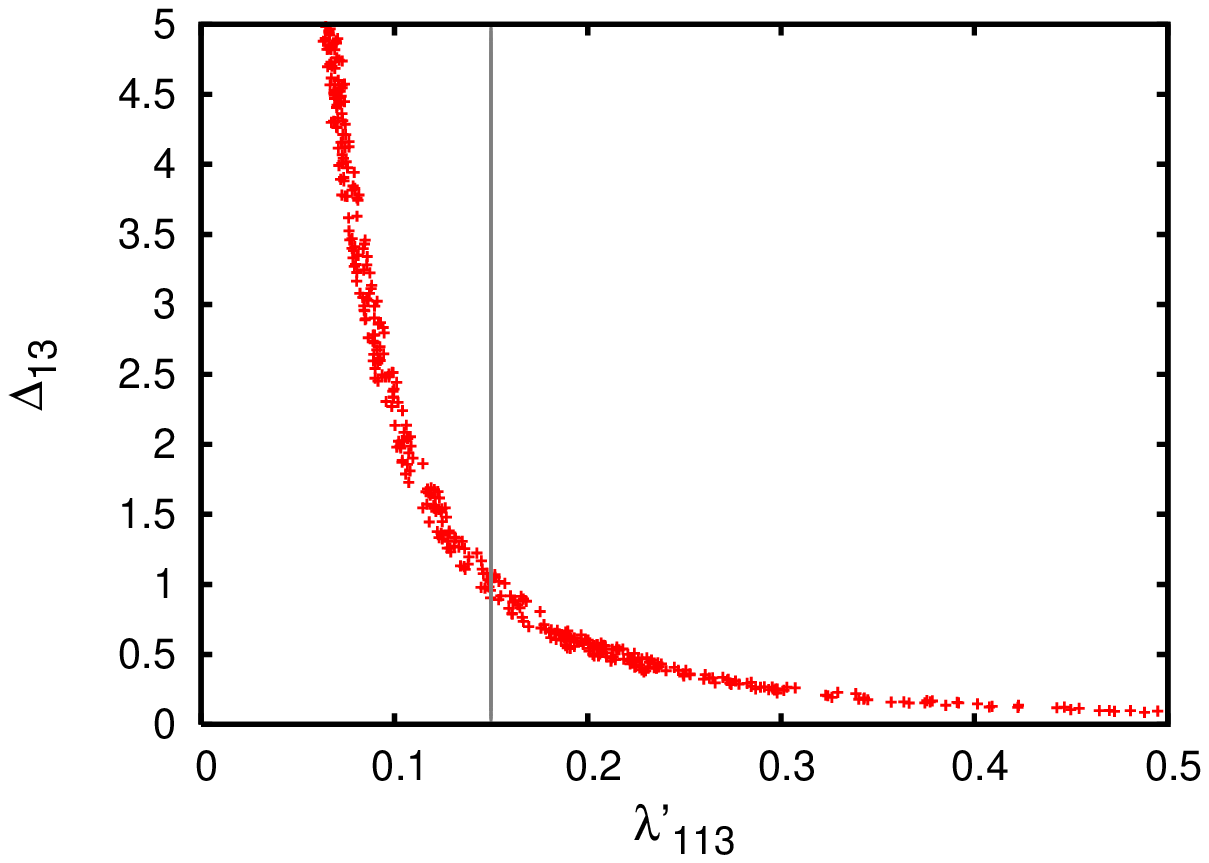,width=8cm,angle=0}}
\parbox{8cm}{\epsfig{file=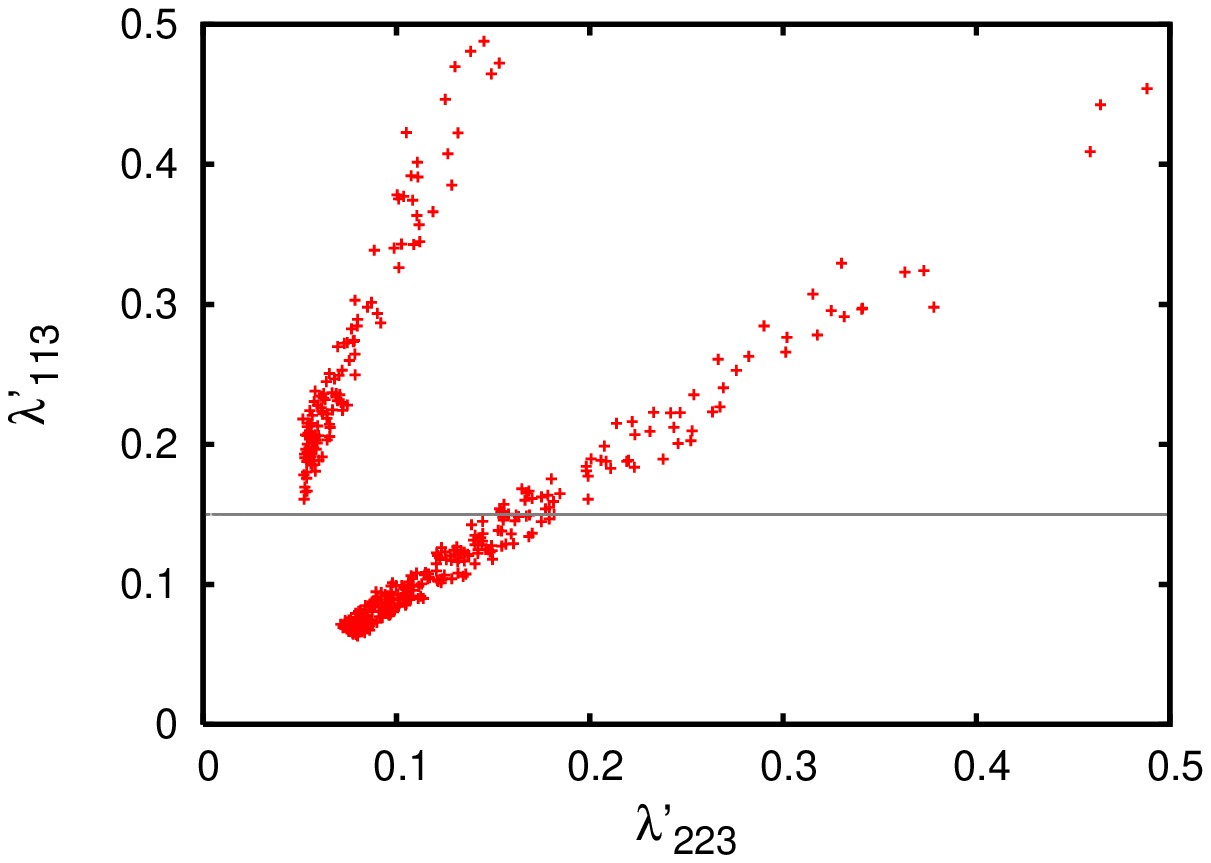,width=8cm,angle=0}}
\parbox{8cm}{\epsfig{file=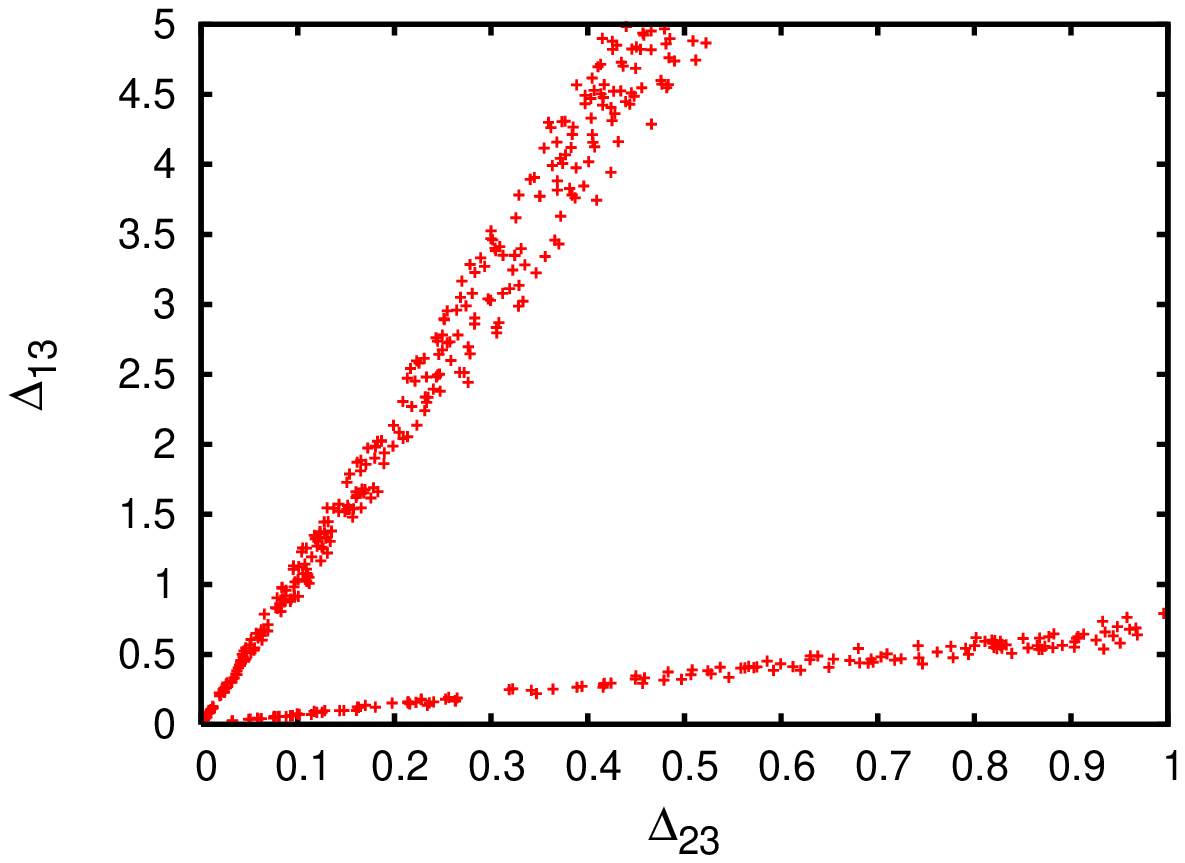,width=8cm,angle=0}}
\caption{Correlation plots for NH case with $\theta_{13} = 0$. The vertical 
  (horizontal) line in the top right (bottom left) panel
  corresponds to the 99\% CL upper limit on $\lambda'_{113}$ (see text).}
\label{nh}
\end{figure}

\begin{figure}[htbp]
\parbox{8cm}{\epsfig{file=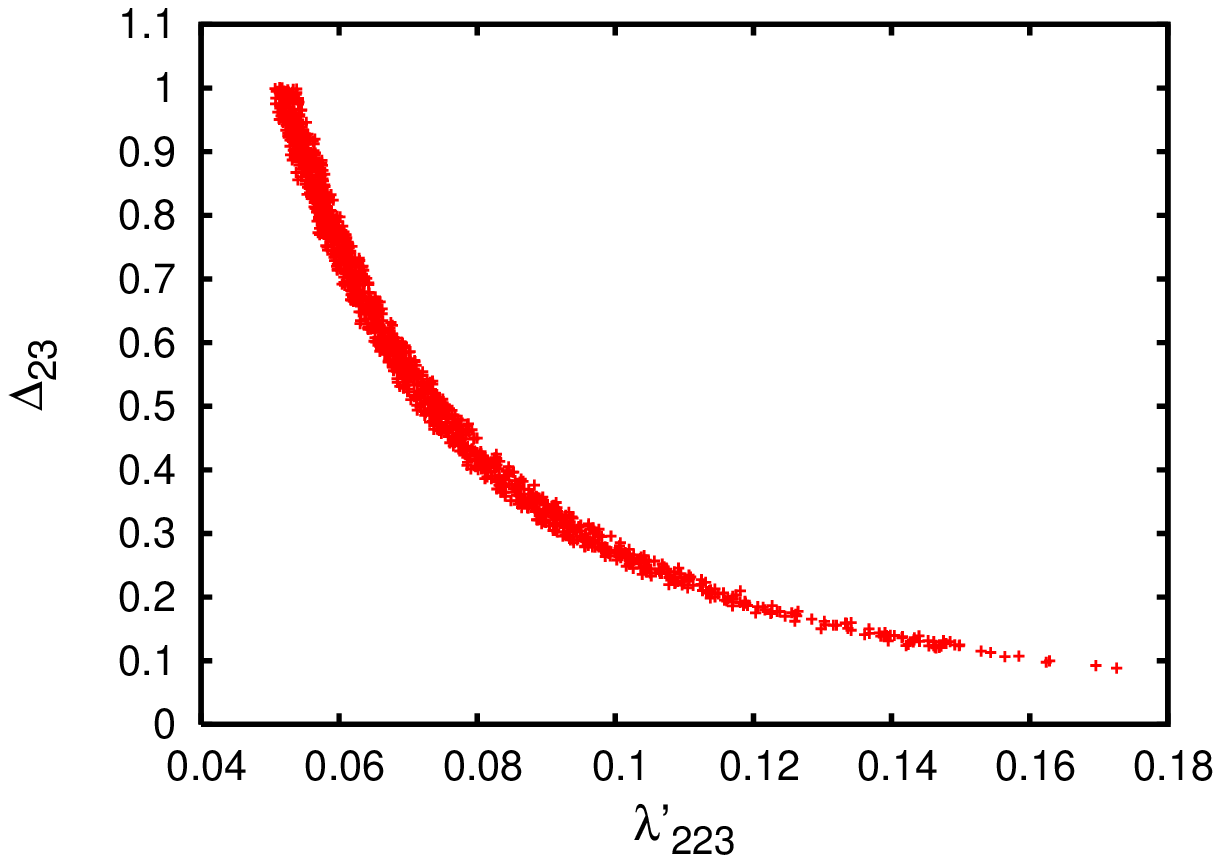,width=8cm,angle=0}}
\parbox{8cm}{\epsfig{file=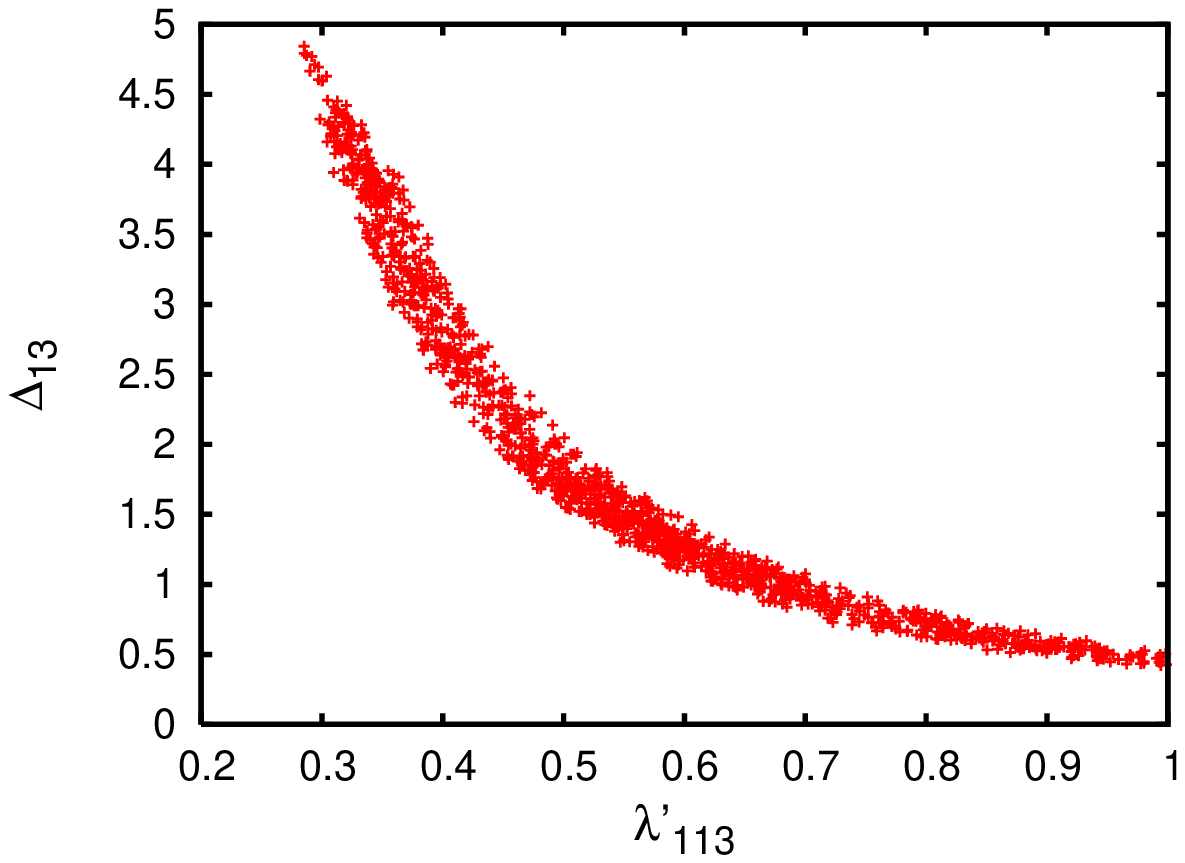,width=8cm,angle=0}}
\parbox{8cm}{\epsfig{file=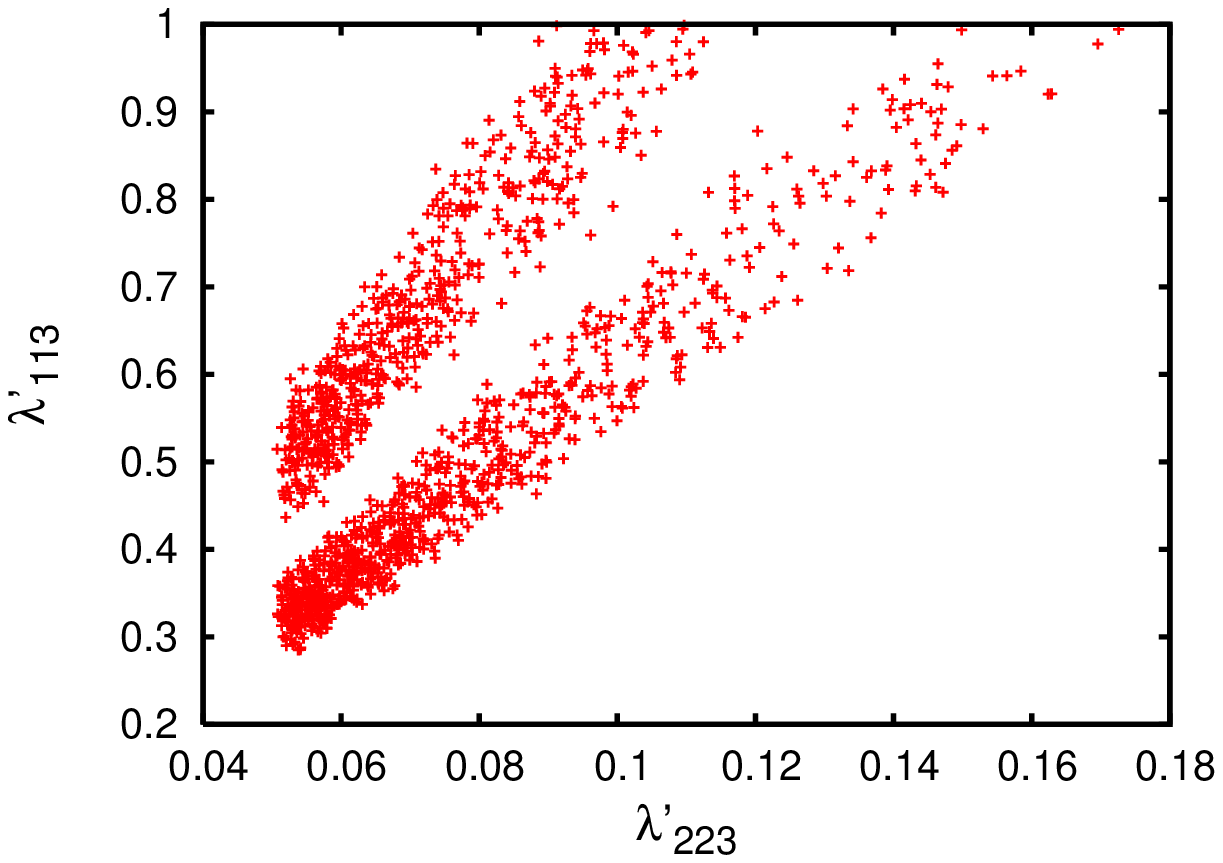,width=8cm,angle=0}}
\parbox{8cm}{\epsfig{file=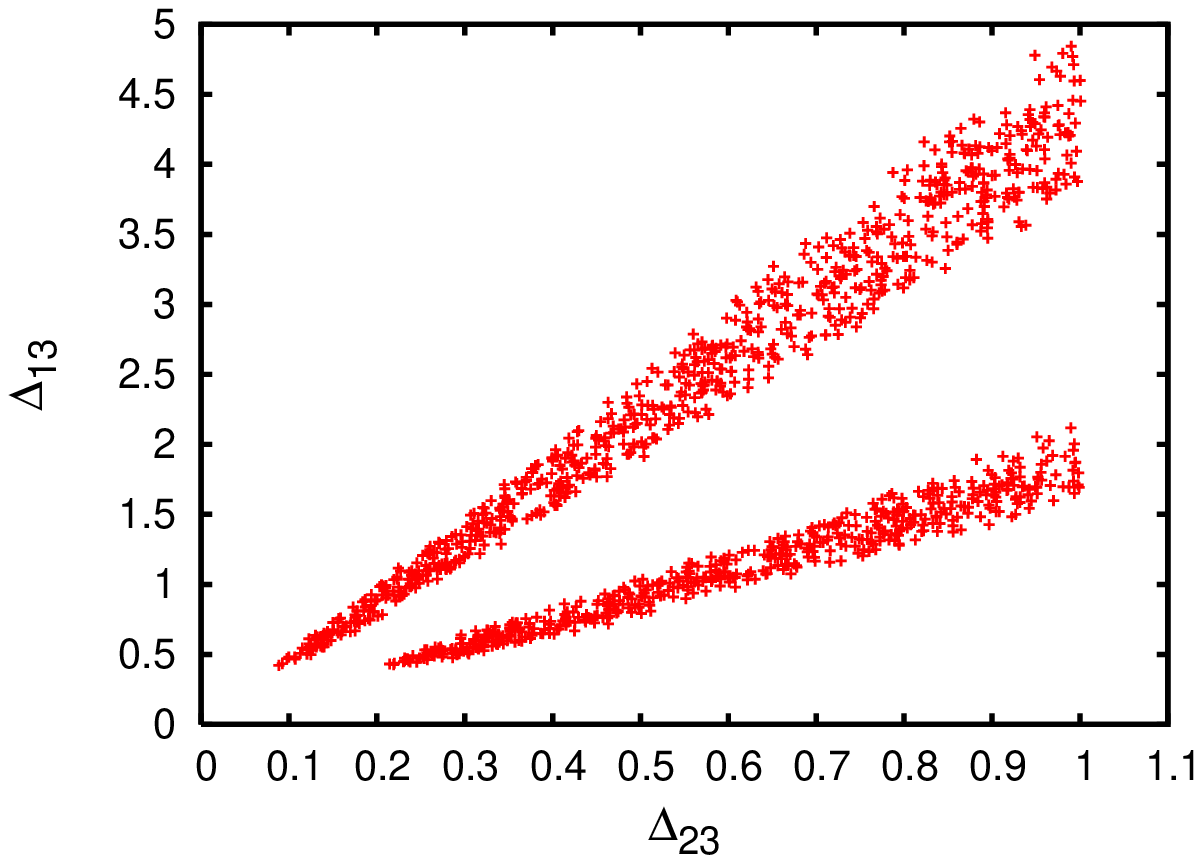,width=8cm,angle=0}}
\caption{Same as in Figure 4 but for the IH scenario.}
\label{ih}
\end{figure}

\begin{figure}[htbp]
\parbox{8cm}{\epsfig{file=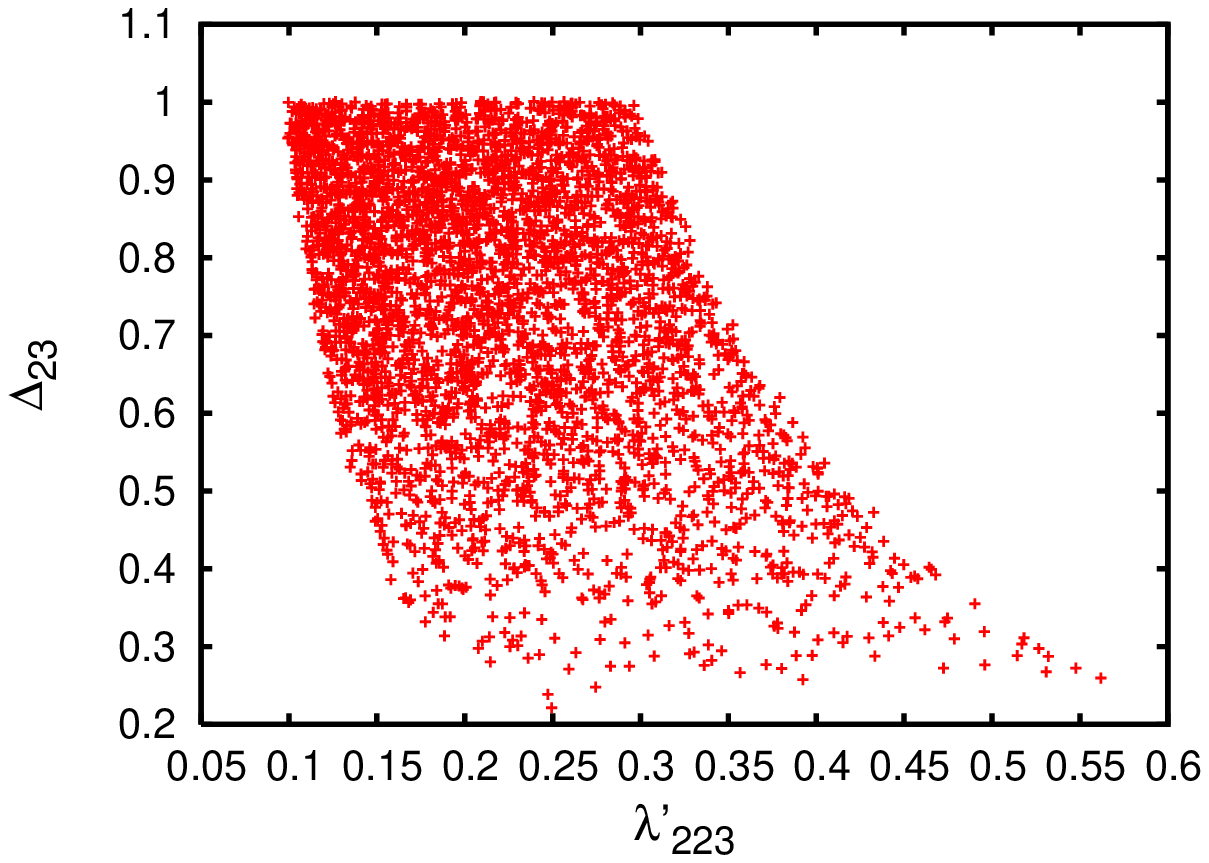,width=8cm,angle=0}}
\parbox{8cm}{\epsfig{file=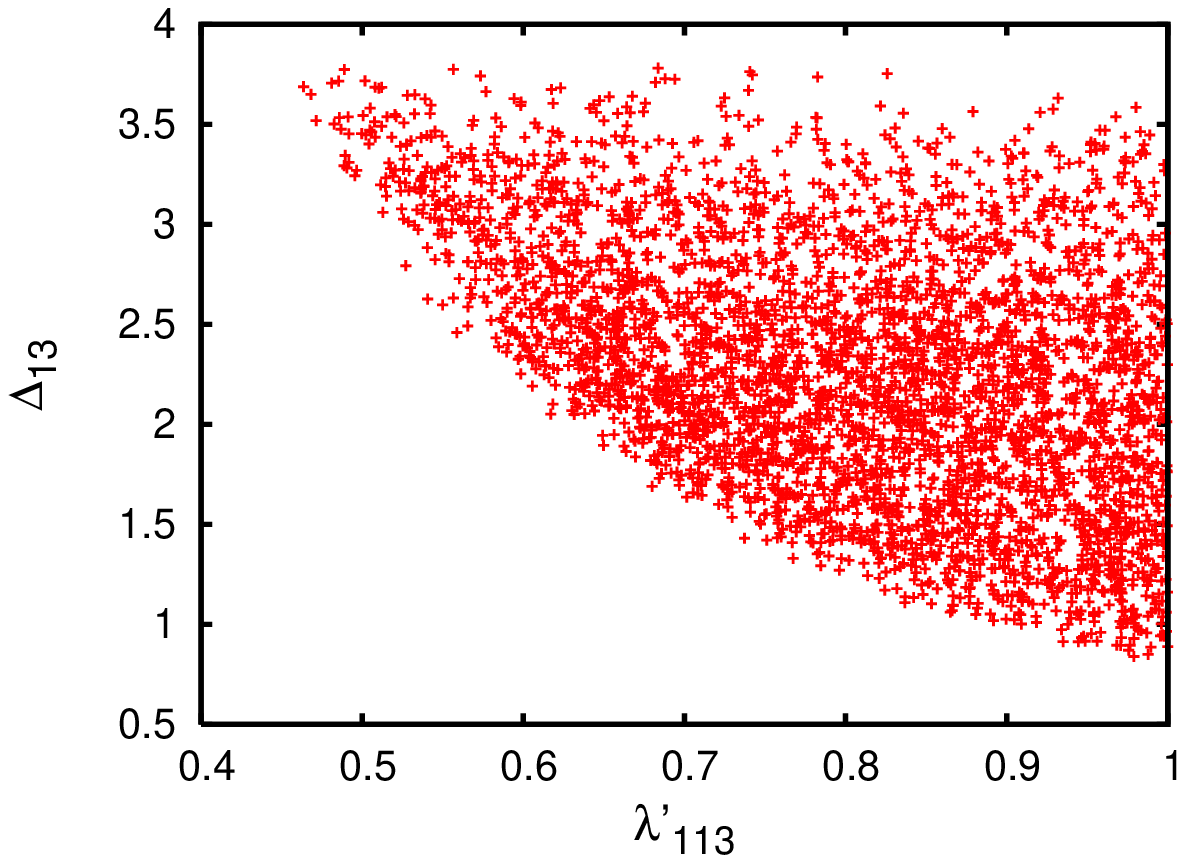,width=8cm,angle=0}}
\parbox{8cm}{\epsfig{file=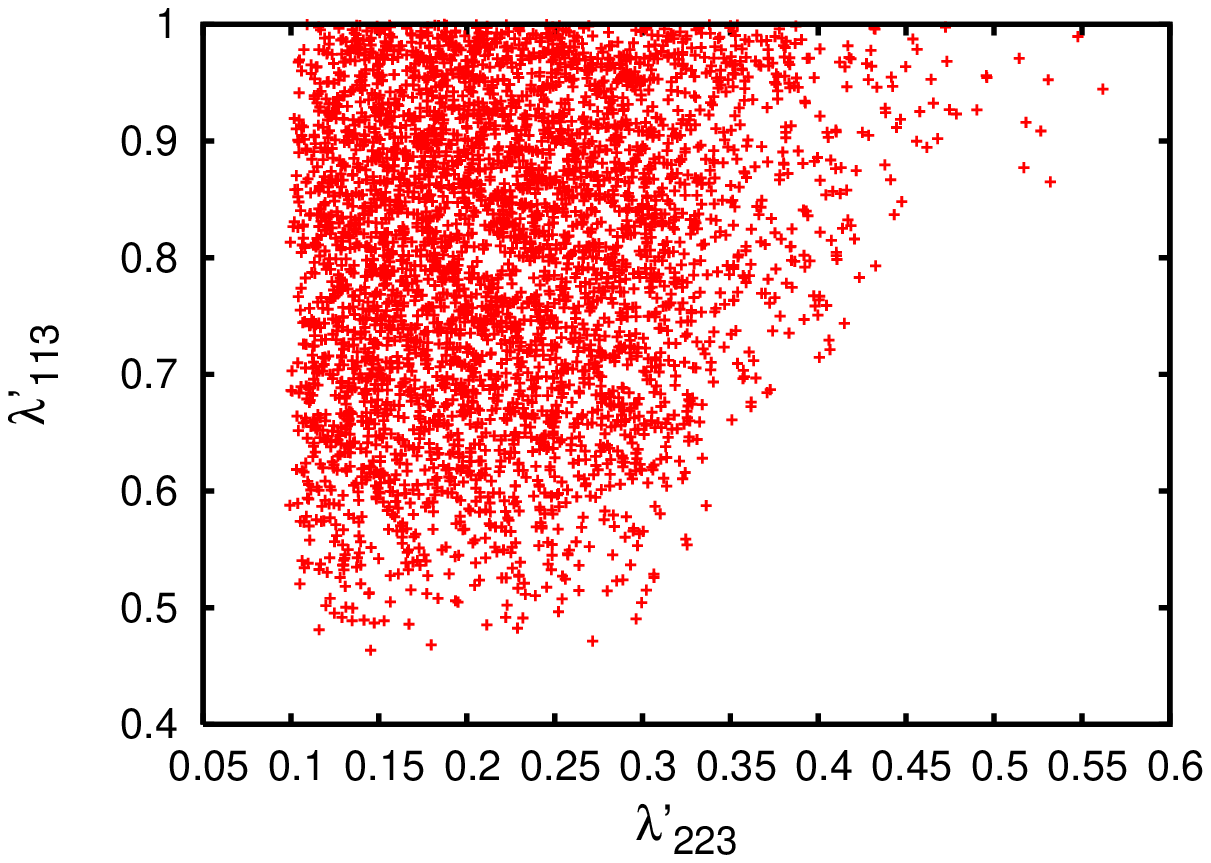,width=8cm,angle=0}}
\parbox{8cm}{\epsfig{file=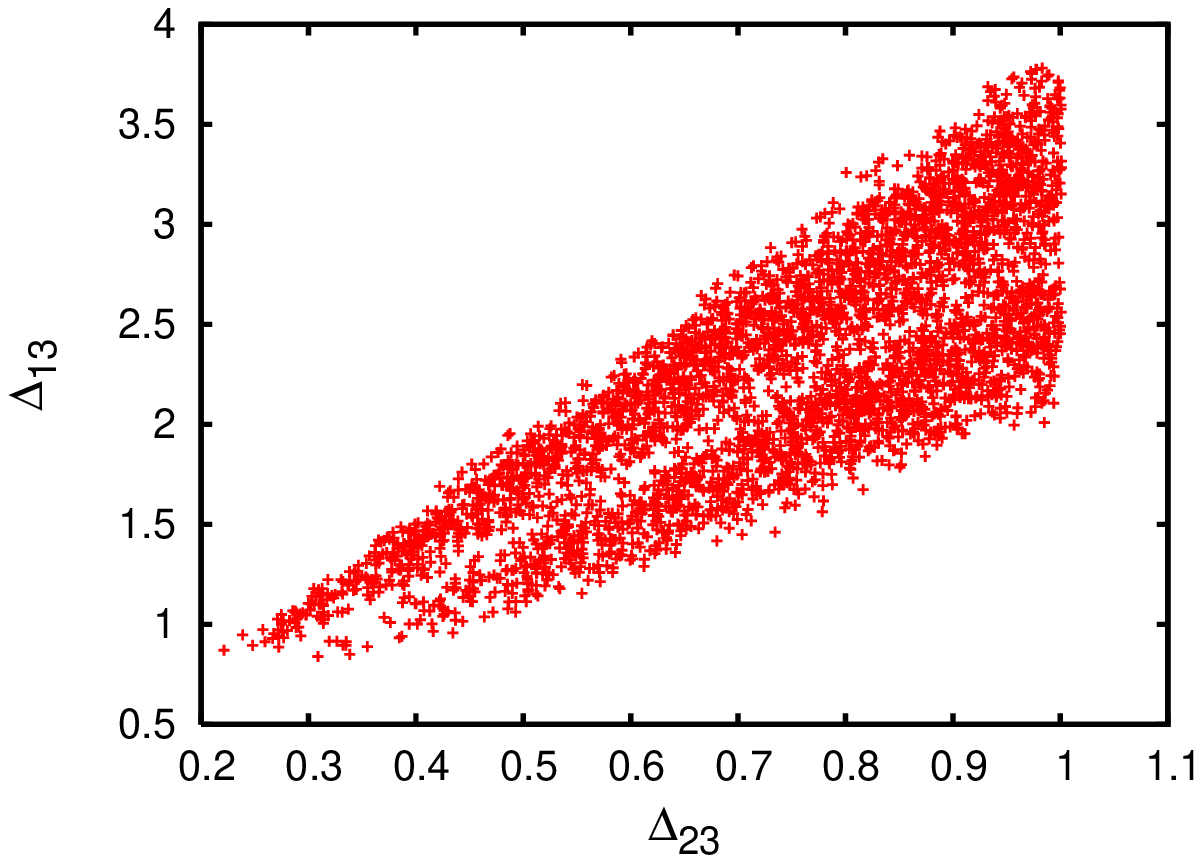,width=8cm,angle=0}}
\caption{Same as in Figure 4 but for the DN scenario.}
\label{dg}
\end{figure}

\begin{figure}[htbp]
\parbox{8cm}{\epsfig{file=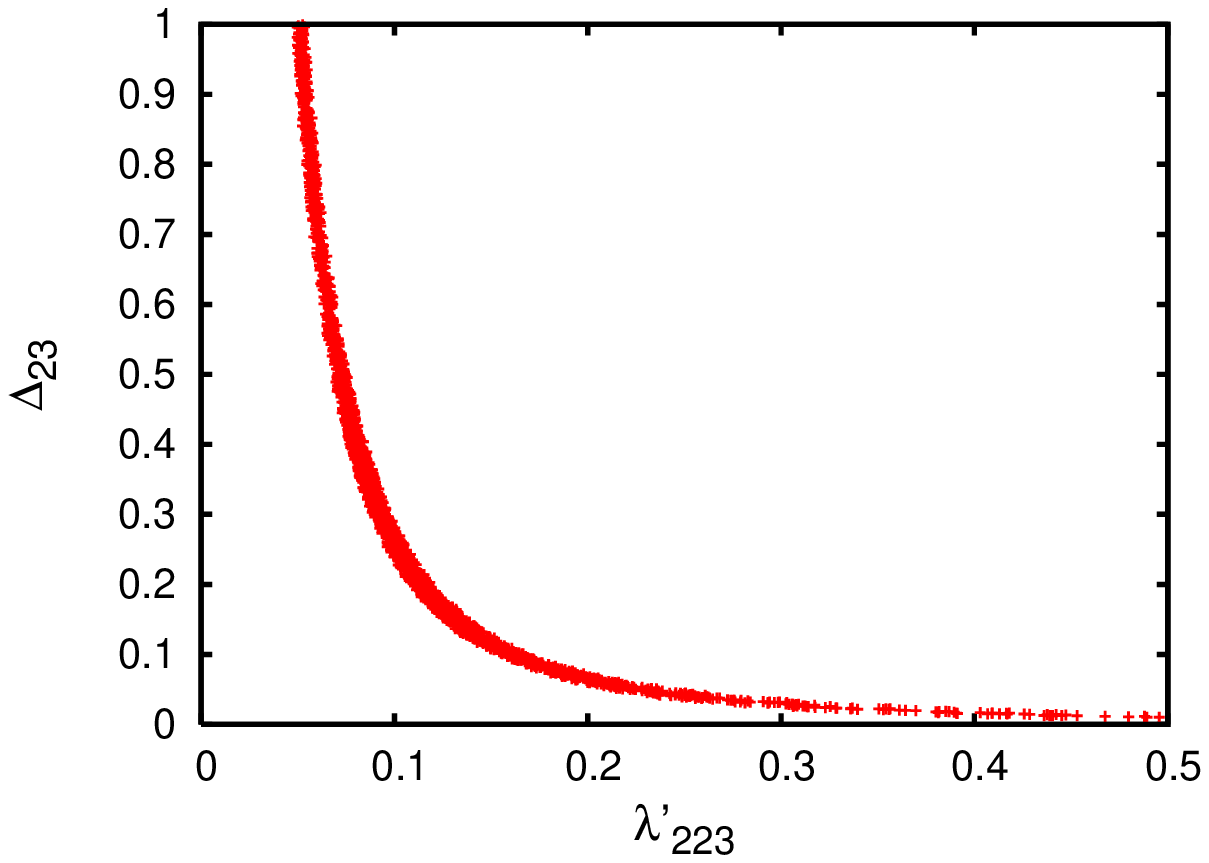,width=8cm,angle=0}}
\parbox{8cm}{\epsfig{file=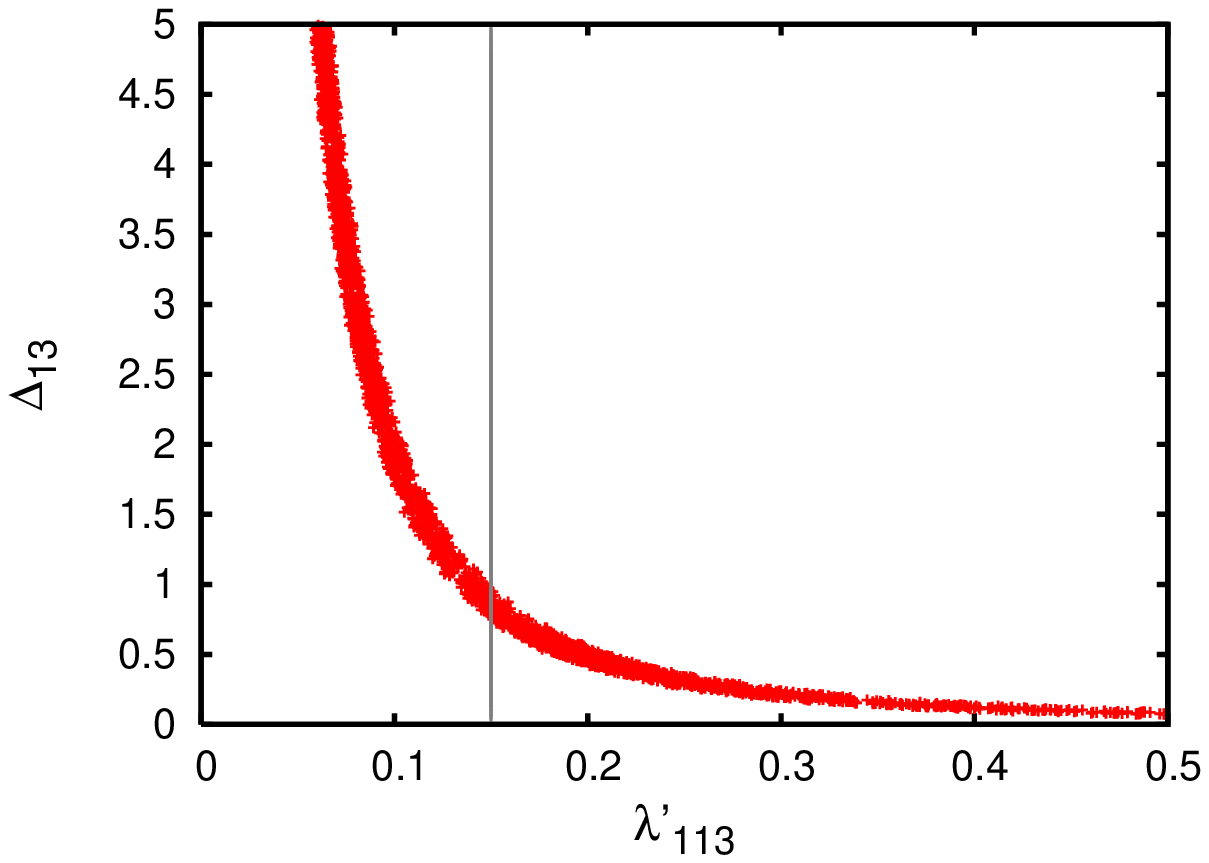,width=8cm,angle=0}}
\parbox{8cm}{\epsfig{file=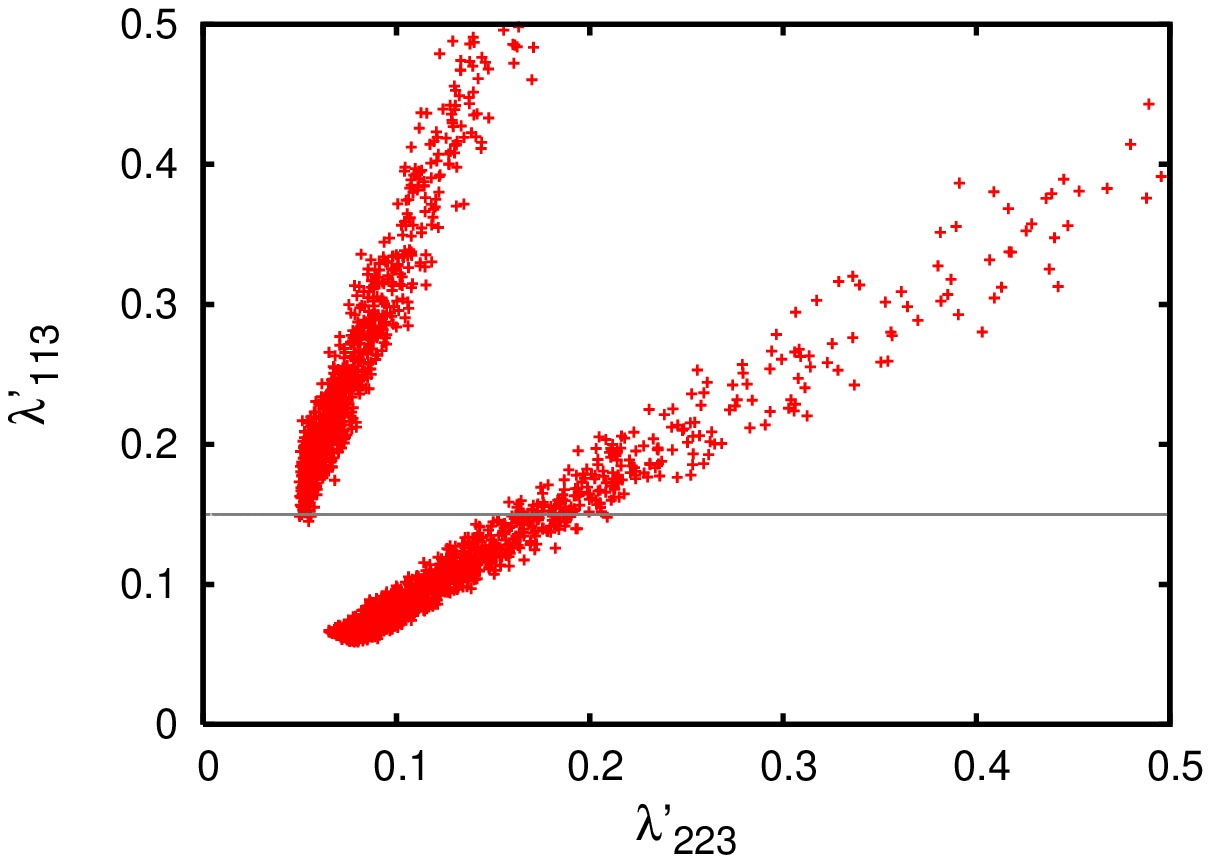,width=8cm,angle=0}}
\parbox{8cm}{\epsfig{file=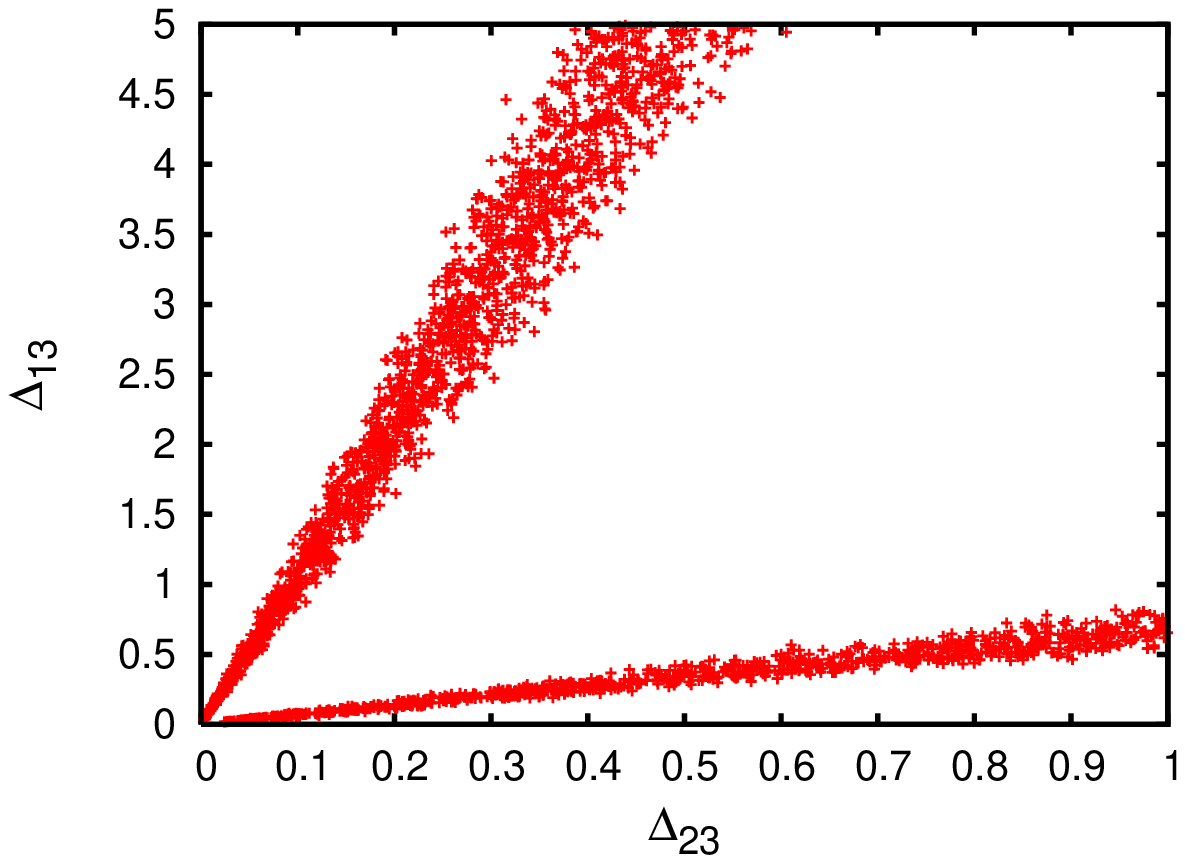,width=8cm,angle=0}}
\caption{Same as in Figure 4 but for $\theta_{13} = 10^{\circ}$. }
\label{nhp}
\end{figure}

As a result of varying all the parameters, there 
are six possible projections of the four-dimensional
scatter plot. In figures \ref{nh}-\ref{nhp}, we show four of them, the other
two not giving any independent information.

The plots are drawn for (i) NH (fig.\ \ref{nh}), (ii) IH (fig.\ \ref{ih}),
(iii) DN (fig.\ \ref{dg}), all with $\theta_{13}=0$, and (iv) NH 
(fig.\ \ref{nhp}), with $\theta_{13}=10^\circ$. 
No such figures are separately shown for the IH and
the DN cases, because (a) there is no appreciable difference with the
corresponding $\theta_{13}=0$ case, and (b) these scenarios are in general
disfavoured by the constraints on $\lambda'_{113}$.

The scatter plots are essentially based on the fact that, 
corresponding to any value of one of the four
aforementioned parameters, we get confined to rather narrow 
intervals of the remaining three, in order to satisfy the relative
values of the neutrino mass matrix elements answering to the tri-bimaximal
mixing pattern. Thus the scatter plots turn into correlation curves (bands)
whose widths come largely from the uncertainties of the neutrino oscillation
data and marginally from the uncertainties in the CKM elements.

To see the most important conclusions, 
let us first concentrate on fig. \ref{nh}. The upper panels show the allowed
regions for $\lambda'$ versus $\Delta$; one goes up as the other goes down.
Qualitatively, this can be understood from the expressions of the $M_\nu$
elements as given in the appendix: the product of the type 
$\lambda'\lambda'\Delta$ appears in the leading contributions. The lower 
left-hand panel shows the correlation between $\lambda'_{113}$ and $\lambda'_
{223}$; taken in conjunction with the upper panels, this also tells the
allowed regions of the corresponding $\Delta$s, and this fact has been 
confirmed in the lower right-hand panel. 
The upper bound on $\lambda'_{113}$, shown by a vertical line in the 
upper right-hand and by a horizontal line in the lower left-hand panels, 
corresponds to the 99\% confidence level limit from charged current 
universality.

It should be noted that the allowed regions for $\lambda'_{113}$ fall
outside this limit for both IH and DN cases. 
However, such constraint, as listed in existing literature,
assumes the existence of no $\lambda$-type couplings, which can invalidate the
bound but play an ineffective role in neutrino mass generation, giving
contributions suppressed by light lepton masses. With these in view, we have
allowed $\lambda'_{113}$ to have values larger than the upper bound found in
the literature, with the caveat that the large values may indicate the
existence of additional interactions of $\lambda$-type.
 
The graphs clearly show that the NH scenario favours larger values of
$\lptwtwth$ than IH, while for $\lambda'_{113}$ it is the other way around.
This is because, in the IH case, one requires the $(1,1)$ element of $M_\nu$
to be of higher magnitude, and one is at a relative disadvantage in the loop
contributions, since the contribution to this element is suppressed by the
down quark mass.  One also gets restricted to rather small values of
$\lptwtwth$ is this case. For the degenerate neutrino case, too,
$\lambda'_{113}$ has to be on the higher side, since the corresponding
contributions do not get the advantage of heavier quark masses. This
re-iterates the difficulty in reconciling the IH and DN scenarios with the
constraints on $\lambda'_{113}$, which can be bypassed through, for example,
the occurrence of additional $R_p$ violating interactions.

While it is true  that the preference of NH over IH results from
the way we have selected our parameters, it should be also be noted
that it is {\it more the result of selecting three $\lambda'$-type
couplings in the range of 0.1 and two squark flavour-violating 
parameters $\Delta$.} It is of course true that one can fit
the IH and DN scenarios with a larger set of R-parity violating
interactions. However, with the so-called `minimal' choice, the
orders of magnitudes of the loop contributions are
not significantly different, so long as the $\lambda'$-parameters
are in the same range, and the squarks of different flavours (due to
the different indices of these parameters) participating in the loops
are in the same mass range. Naturally, the contributions will be
smaller with more massive squarks; then higher values of the R-parity
violating couplings than what is indicated can be accommodated. Thus,
while it is not our goal to establish the preference of one scenario
over the other,  what we successfully show is that one can generate 
neutrino masses with large R-parity violating couplings, and that a 
pattern follows from a minimal choice, which does not necessarily
depend on which three parameters are involved.

\section{Correlated signals: some speculations}
\label{sec5}
As we have noted earlier, the small values of $\lambda'_{223}$ and $\lambda'_
{323}$ can enhance the $D_s\to \mu(\tau)\nu$ branching ratio marginally.
However, if one indeed entertains the possibility of some other $\lambda$-type
interaction to save the IH or the DN picture, it is possible that these two
couplings may become large. The lepton flavour violating (LFV) decay $\Upsilon
\to\mu\tau$ is, again, only marginally enhanced, and is still well below the
experimental limit. However, a positive signal in this channel would be very
interesting from the neutrino perspective. The same comment applies to other
LFV decays, like $D^0\to e\mu$, driven by $\lambda'_{113}\lambda'_{223}$.

One of the most interesting low-energy effects for this scenario is the change
in the branching ratio of $K^+\to\pi^+\nu\bar{\nu}$. The decay, based on $s\to
d\nu\bar{\nu}$, is again controlled by $\lambda'_{113}\lambda'_{223}$.  The
experimental number is $B(K^+ \longrightarrow \pi^+ \nu {\bar{\nu}}) =
(1.47^{+1.30}_{-0.89}) \times 10^{-10}$ \cite{pdg}, while the SM prediction is
about $(0.8\pm 0.1)\times 10^{-10}$. It was pointed out in \cite{welzel} that
an exact upper bound is difficult to obtain considering the interplay of the
SM, the $R_p$ conserving SUSY and the $R_p$ violating SUSY, but it can safely
be said that with couplings of the order that we have used in this work, the
$R_p$ violating amplitude may even be larger than the SM amplitude. In that
case, this mode cannot be used as a clean channel for extracting
$\sin(2\beta)$. Measurement of the said angle and a comparison with the
charmonium result will again be crucial for our ansatz.

\section{Summary and conclusions}
\label{sec6}
We have considered scenarios where $R_p$ violating couplings can be large, and
the neutrino mass matrix can still be generated in a manner consistent with
observed results. This, we argue, can be possible if there are only a few
couplings of this type, so that the combinations necessary for one-loop
neutrino masses are not available. Two-loop contributions come to one's
advantage in such situations, together with the possibility of flavour
violation in the sfermion mass matrices. Considering the $\lambda'$-type
couplings, we have demonstrated this; with three such couplings and two squark
flavour violating parameters, the NH scenario can be reproduced, guiding one
to a specific region of the parameter space. For the IH and DN cases, however,
this requires the value of at least one coupling to come into conflict with
observable constraints unless one postulates additional R-parity violating
terms in the superpotential. Of course, there may be more than one choice of
the set of R-parity violating couplings leading to two-loop neutrino masses,
and the exact numerical consequences in the neutrino sector can be dependent
on which  $\lambda'$-couplings actually exist.

Similar conclusions can be established if one includes the bilinear R-parity
violating terms in the superpotential. One neutrino state acquires a
tree-level mass in such a case, thus relaxing the constraint that seems to
loom large on the parameter $\lambda'_{113}$ as discussed above. The two
remaining couplings (with values in the range 0.1-1.0) and the squark flavour
violation parameters can then generate the remaining terms in the mass matrix
at the two-loop level. This may make the IH and DN cases less constrained.

In conclusion, large trilinear R-parity violating interactions are not
necessarily an impediment to the explanation of neutrino masses and mixing.
Thus if some phenomenon observed in the laboratory points strongly towards
such large interaction strength, it may still explain the neutrino sector
perfectly well, provided that only a few R-parity violating interactions occur
in nature with sizable strength.

{\bf ACKNOWLEDGEMENTS}

We thank Abhijit Bandyopadhyay and Sourov Roy for helpful discussions.  AK is
supported by Project No. 2007/37/9/BRNS of DAE, Government of India. The work
of PD, BM and SN was partially supported by funding available from the
Department of Atomic Energy, Government of India, for the Regional Centre for
Accelerator-based Particle Physics, Harish-Chandra Research Institute.

{\bf APPENDIX}

The two-loop matrix elements are expressed in terms of the following
variables.
\begin{eqnarray}
\xi_t = V^*_{ts}V_{tb} \\
\xi'_t = V^*_{td}V_{tb} \\
\xi_c = V^*_{cs}V_{cb} \\
\xi'_c = V^*_{cd}V_{cb} \\
x_i = m^2_i/M^2_W 
\end{eqnarray}
The generic loop-functions with proper arguments are listed below. There are
two types of such functions, depending on whether they depend on lepton masses
or not.

Functions, first set: $i=e,\mu,\tau$.
\begin{eqnarray}
F_1 (x_t,x_W) = \frac{3x_t-1}{4(x_t-1)} - \frac{x^2_t\log
  x_t}{2(x_t-1)^2} 
\end{eqnarray}
\begin{eqnarray} 
F_2 (x_t,x_W) = 1 - \frac{x_t\log x_t}{x_t-1}
\end{eqnarray}
\begin{eqnarray}
F_1 (x_t,x_W) - F_2 (x_t,x_W) =
  \frac{x_t(x_t-2)}{2(x_t-1)^2}\log x_t - \frac{x_t-3}{4(x_t-1)} 
\end{eqnarray}
\begin{eqnarray}
  F_3 (x_t,x_W) = -2 F_1 (x_t,x_W) 
\end{eqnarray}
\begin{eqnarray}
 F_4 (x_c,x_i) = \frac{1-x_c+x_c\log x_c}{1-x_c} + \frac{x_c\log x_c - x_c -
  x_i\log x_i + x_i}{x_c - x_i}
\end{eqnarray}
\begin{eqnarray}
 F_5 (x_c,x_i) &=& -\frac{1}{(x_c-x_W)^2} \Big[ \frac{1}{2}x^2_c\log x_c -
  \frac{1}{4}x^2_c - x_wx_c\log x_c + x_wx_c \nonumber\\ &&+
  \frac{1}{2}x^2_W\log x_W - \frac{3}{4}x^2_W \Big] + (x_W \leftrightarrow
  x_i)
\end{eqnarray}
\begin{eqnarray}
 F_6 (x_c,x_i) = \frac{1}{(x_w-x_c)} \Big[ \frac{1}{2}x^2_w\log x_w -
  \frac{1}{4}x^2_w - \frac{1}{2}x^2_c\log x_c - \frac{1}{4}x^2_c \Big]
  \nonumber\\ - (x_W \leftrightarrow x_i)
\end{eqnarray}
\begin{eqnarray}
 F_7 (x_c,x_i) &=& \frac{1}{(x_w-x_c)^3} \Big[ \big(\frac{1}{3}x^3_w -
 x^2_wx_c + x_wx^2_c\big)\log x_w \nonumber\\ && - \big(\frac{1}{9}x^3_w -
 \frac{1}{2}x^2_wx_c + x_wx^2_c\big) - \frac{1}{3}x^3_c \log x_c -
 \frac{11}{18} x^3_c \Big] - (x_W \leftrightarrow x_i)
\end{eqnarray}
Functions, second set: 
\begin{eqnarray}
F_8 (x_{\tilde q}, x_b) = \frac{1}{x_{\tilde q} - x_b} - \frac{x_b(\log
x_{\tilde q} - \log x_b)}{(x_{\tilde q} - x_b)^2}
\end{eqnarray}
\begin{eqnarray}
F_9 (x_{\tilde q}, x_b) = \frac{4}{(x_{\tilde q} - x_b)^2} \left[ \frac{1}{2}
  x^2_{\tilde q} \log x_{\tilde q} - \frac{1}{4} x^2_{\tilde q} - x_{\tilde q}
  x_b \log x_{\tilde q} + x_{\tilde q} x_b + \frac{1}{2} x^2_b \log x_b -
  \frac{3}{4}x^2_b \right]
\end{eqnarray}
\begin{eqnarray}
F_{10} (x_{\tilde q}, x_b, x_s) = F_9 (x_{\tilde q}, x_b) - F_9 (x_{\tilde q},
x_s)
\end{eqnarray}
\begin{eqnarray}
F_{11} (x_{\tilde q}, x_b, x_s) = F_6 (x_{\tilde q}, x_b) - F_6
(x_{\tilde q}, x_s)
\end{eqnarray}
\begin{eqnarray}
F_{12} (x_{\tilde q}, x_b) = \frac{\log x_{\tilde q} - \log x_b}{x_{\tilde q}
  - x_b}
\end{eqnarray}
\begin{eqnarray}
F_{13} (x_{\tilde q}, x_b) = \frac{1}{(x_{\tilde q}-x_b)} \Big[
  \frac{1}{2}x^2_{\tilde q} \log x_{\tilde q} - \frac{1}{4}x^2_{\tilde q} -
  \frac{1}{2}x^2_b\log x_b + \frac{1}{4}x^2_b \Big]
\end{eqnarray}

\underline{Matrix element $M_{\nu}(1,1)$}: 

$M_{\nu}(1,1)$: Diagram 2(a) with $\phi$
\begin{eqnarray}
&& \lambda'_{113}\lambda'_{113}~ \frac{g^2}{4M^2_W}~ \frac{1}{(16\pi^2)^2}
~\xi'_t ~M_{susy} ~\frac{m_bm_d}{m^2_d-m^2_b} ~\tilde{\Delta}_{13} \Big[
m^2_t(F_1-F_2)F_{10} (x_{\tilde q}, x_b, x_d) \nonumber\\ &&- m^2_t x_d F_2
F_{11} (x_{\tilde q}, x_b, x_d) + m^2_d F_1 F_{10} (x_{\tilde q}, x_b, x_d)
\Big] 
\end{eqnarray}
$M_{\nu}(1,1)$: Diagram 2(a) with $W$
\begin{eqnarray}
&& \lambda'_{113}\lambda'_{113}~ \frac{g^2}{4}~ \frac{1}{(16\pi^2)^2}
~M_{susy} ~\frac{m_bm_d}{m^2_d-m^2_b} ~\tilde{\Delta}_{13} \Big[ \xi'_t [F_3
(x_t,x_W)-F_3 (x_u,x_W)] \nonumber\\ && + \xi'_c [F_3 (x_c,x_W)-F_3 (x_u,x_W)]
\Big] F_{10} (x_{\tilde q}, x_b, x_d) 
\end{eqnarray}
$M_{\nu}(1,1)$: Diagram 2(b) with $\phi$
\begin{eqnarray}
&& \lambda'_{113}\lambda'_{113}~ \frac{g^2}{4M^2_W}~ \frac{1}{(16\pi^2)^2}
~V_{ub} ~M_{susy} ~{m_bm_d} ~\tilde{\Delta}_{13} \Big[
\frac{m^2_e}{M^2_W-m^2_e} [x_uF_4 (x_u,x_e)F_8 (x_{\tilde q}, x_b) \nonumber\\
&&- F_4 (x_u,x_e)F_9 (x_{\tilde q}, x_b) - F_5 (x_u,x_e)F_9 (x_{\tilde q},
x_b)] \Big] 
\end{eqnarray}
$M_{\nu}(1,1)$: Diagram 2(c) with $\phi$
\begin{eqnarray}
&& \lambda'_{113}\lambda'_{113}~ g^2 \frac{1}{(16\pi^2)^2} ~V^*_{ud} \Big[
\frac{x_u - x_d}{x_w - x_e} m_{e} F_5 (x_u,x_e) F_{13} (x_{\tilde q},x_d)
\nonumber\\ && + \frac{x_d}{x_w - x_e} m_{e} [F_7 (x_u,x_e) F_{13} (x_{\tilde
q},x_d) - F_6 (x_u,x_e) F_{12} (x_{\tilde q},x_d)] \Big] 
\end{eqnarray}

\underline{Matrix element $M_{\nu}(2,2)$}: 

$M_{\nu}(2,2)$: Diagram 2(a) with $\phi$
\begin{eqnarray}
&& \lambda'_{223}\lambda'_{223}~ \frac{g^2}{4M^2_W}~ \frac{1}{(16\pi^2)^2}
~\xi_t ~M_{susy} ~\frac{m_bm_s}{m^2_s-m^2_b} ~\tilde{\Delta}_{23} \Big[
m^2_t(F_1-F_2)F_{10} (x_{\tilde q}, x_b, x_s) \nonumber\\ && - m^2_t x_s F_2
F_{11} (x_{\tilde q}, x_b, x_s) + m^2_s F_1 F_{10} (x_{\tilde q}, x_b, x_s)
\Big] 
\end{eqnarray}
$M_{\nu}(2,2)$: Diagram 2(a) with $W$
\begin{eqnarray}
&& \lambda'_{223}\lambda'_{223}~ \frac{g^2}{4}~ \frac{1}{(16\pi^2)^2}
~M_{susy} ~\frac{m_bm_s}{m^2_s-m^2_b} ~\tilde{\Delta}_{23} \Big[ \xi_t [F_3
(x_t,x_W)-F_3 (x_u,x_W)] \nonumber\\ &&+ \xi_c [F_3 (x_c,x_W)-F_3 (x_u,x_W)]
\Big] F_{10} (x_{\tilde q}, x_b, x_s) 
\end{eqnarray}
$M_{\nu}(2,2)$: Diagram 2(b) with $\phi$
\begin{eqnarray}
&& \lambda'_{223}\lambda'_{223}~ \frac{g^2}{4M^2_W}~ \frac{1}{(16\pi^2)^2}
~V_{cb} ~M_{susy} ~{m_bm_s} ~\tilde{\Delta}_{23} \Big[
\frac{m^2_\mu}{M^2_W-m^2_\mu} [x_cF_4 (x_c,x_\mu)F_6 (x_{\tilde q}, x_b)
\nonumber\\ && - F_4 (x_c,x_\mu)F_9 (x_{\tilde q}, x_b) - F_5 (x_c,x_\mu)F_9
(x_{\tilde q}, x_b)] \Big] 
\end{eqnarray}
$M_{\nu}(2,2)$: Diagram 2(c) with $\phi$
\begin{eqnarray}
&& \lambda'_{223}\lambda'_{223}~ g^2 \frac{1}{(16\pi^2)^2} ~V^*_{cs} \Big[
\frac{x_c - x_s}{x_w - x_\mu} m_{\mu} F_5 (x_c,x_\mu) F_{13} (x_{\tilde
q},x_s) \nonumber\\ && + \frac{x_s}{x_w - x_\mu} m_{\mu} [F_7 (x_c,x_\mu)
F_{13} (x_{\tilde q},x_s) - F_6 (x_c,x_\mu) F_{12} (x_{\tilde q},x_s)] \Big]
\end{eqnarray}
Matrix element $M_{\nu}(3,3)$ same as $M_{\nu}(2,2)$ with $\mu$ replaced by
$\tau$. \\

\underline{Matrix element $M_{\nu}(1,2) = M_{\nu}(2,1)$}: 

$M_{\nu}(1,2)$: Diagram 2(a) with $\phi$
\begin{eqnarray}
&& \lambda'_{113}\lambda'_{223}~ \frac{g^2}{8M^2_W}~ \frac{1}{(16\pi^2)^2}
~M_{susy} \Big( \Big[ \xi_t ~\frac{m_bm_d}{m^2_s-m^2_b} ~\tilde{\Delta}_{13}
\Big] \Big[ m^2_t(F_1-F_2)F_{10} (x_{\tilde q}, x_b, x_s) \nonumber \\ && -
m^2_t x_s F_2 F_{11} (x_{\tilde q}, x_b, x_s) + m^2_s F_1 F_{10} (x_{\tilde
q}, x_b, x_s) \Big] \nonumber\\ && + \Big[ \xi'_t ~\frac{m_bm_s}{m^2_d-m^2_b}
~\tilde{\Delta}_{23} \Big] \Big[ m^2_t(F_1-F_2)F_{10} (x_{\tilde q}, x_b, x_d)
\nonumber\\ && - m^2_t x_d F_2 F_{11} (x_{\tilde q}, x_b, x_d) + m^2_d F_1
F_{10} (x_{\tilde q}, x_b, x_d) \Big] \Big)
\end{eqnarray}
$M_{\nu}(1,2)$: Diagram 2(a) with $W$
\begin{eqnarray}
&& \lambda'_{113}\lambda'_{223}~ \frac{g^2}{8}~ \frac{1}{(16\pi^2)^2}
~M_{susy} \Big( \Big[ \frac{m_bm_d}{m^2_s-m^2_b} ~\tilde{\Delta}_{13} \Big]
\Big[ \xi_t [F_3 (x_t,x_W)-F_3 (x_u,x_W)] \nonumber \\ && + \xi_c [F_3
(x_c,x_W)-F_3 (x_u,x_W)] \Big] F_{10} (x_{\tilde q}, x_b, x_s) + \Big[
\frac{m_bm_s}{m^2_d-m^2_b} ~\tilde{\Delta}_{23} \Big] \Big[ \xi'_t [F_3
(x_t,x_W)\nonumber \\ && -F_3 (x_u,x_W)] + \xi'_c [F_3 (x_c,x_W)-F_3
(x_u,x_W)] \Big] F_{10} (x_{\tilde q}, x_b, x_d) \Big)
\end{eqnarray}
$M_{\nu}(1,2)$: Diagram 2(b) with $\phi$
\begin{eqnarray}
&& \lambda'_{113}\lambda'_{223}~ \frac{g^2}{8M^2_W}~ \frac{1}{(16\pi^2)^2}
~M_{susy} \Big( \Big[ V_{cb}~{m_bm_d}~\tilde{\Delta}_{13} \Big] \Big[
\frac{m^2_\mu}{M^2_W-m^2_\mu} [x_cF_4 (x_c,x_\mu)F_8 (x_{\tilde q}, x_b)
\nonumber \\ && - F_4 (x_c,x_\mu)F_9 (x_{\tilde q}, x_b) - F_5 (x_c,x_\mu)F_9
(x_{\tilde q}, x_b)] \Big] + \Big[ V_{ub} ~{m_bm_s} ~\tilde{\Delta}_{23} \Big]
\nonumber \\ && \Big[ \frac{m^2_e}{M^2_W-m^2_e} [x_uF_4 (x_u,x_e)F_8
(x_{\tilde q}, x_b) - F_4 (x_u,x_e)F_9 (x_{\tilde q}, x_b) - F_5 (x_u,x_e)F_9
(x_{\tilde q}, x_b)] \Big] \Big)
\end{eqnarray}
$M_{\nu}(1,2)$: Diagram 2(c) with $\phi$
\begin{eqnarray}
&& \lambda'_{113}\lambda'_{223}~ g^2 \frac{1}{(16\pi^2)^2} \Big[ V^*_{cs}
\Big( \frac{x_c - x_s}{x_w - x_\mu} m_{\mu} F_5 (x_c,x_\mu) F_{13} (x_{\tilde
q},x_s) \nonumber\\ && + \frac{x_s}{x_w - x_\mu} m_{\mu} [F_7 (x_c,x_\mu)
F_{13} (x_{\tilde q},x_s) - F_6 (x_c,x_\mu) F_{12} (x_{\tilde q},x_s)] \Big) +
\nonumber\\ && V^*_{ud} \Big( \frac{x_u - x_d}{x_w - x_e} m_{e} F_5 (x_u,x_e)
F_{13} (x_{\tilde q},x_d) \nonumber\\ && + \frac{x_d}{x_w - x_e} m_{e} [F_7
(x_u,x_e) F_{13} (x_{\tilde q},x_d) - F_6 (x_u,x_e) F_{12} (x_{\tilde q},x_d)]
\Big) \Big]
\end{eqnarray}
Matrix elements $M_{\nu}(1,3)$ and $M_{\nu}(3,1)$ are same as $M_{\nu}(1,2)$
with $\mu$ replaced by $\tau$.

\underline{Matrix element $M_{\nu}(2,3) = M_{\nu}(3,2)$}: 

$M_{\nu}(2,3)$: Diagram 2(a) with $\phi$
\begin{eqnarray}
&& \lambda'_{223}\lambda'_{323}~ \frac{g^2}{8M^2_W}~ \frac{1}{(16\pi^2)^2}
~\xi_t ~M_{susy} ~\frac{m_bm_s}{m^2_s-m^2_b} ~\tilde{\Delta}_{23} \Big[
m^2_t(F_1-F_2)F_{10} (x_{\tilde q}, x_b, x_s) \nonumber\\ && - m^2_t x_s F_2
F_{11} (x_{\tilde q}, x_b, x_s) + m^2_s F_1 F_{10} (x_{\tilde q}, x_b, x_s)
\Big]
\end{eqnarray}
$M_{\nu}(2,3)$: Diagram 2(a) with $W$
\begin{eqnarray}
&& \lambda'_{223}\lambda'_{323}~ \frac{g^2}{8}~ \frac{1}{(16\pi^2)^2}
~M_{susy} ~\frac{m_bm_s}{m^2_s-m^2_b} ~\tilde{\Delta}_{23} \Big[ \xi_t [F_3
(x_t,x_W)-F_3 (x_u,x_W)] \nonumber\\ && + \xi_c [F_3 (x_c,x_W)-F_3 (x_u,x_W)]
\Big] F_{10} (x_{\tilde q}, x_b, x_s) 
\end{eqnarray}
$M_{\nu}(2,3)$: Diagram 2(b) with $\phi$
\begin{eqnarray}
&& \lambda'_{223}\lambda'_{323}~ \frac{g^2}{8M^2_W}~ \frac{1}{(16\pi^2)^2}
~V_{cb} ~M_{susy} ~{m_bm_s} ~\tilde{\Delta}_{23} \Big[
\frac{m^2_\mu}{M^2_W-m^2_\mu}~ [x_cF_4 (x_c,x_\mu)F_8 (x_{\tilde q}, x_b)
\nonumber\\ && - F_4 (x_c,x_\mu)F_9 (x_{\tilde q}, x_b) - F_5 (x_c,x_\mu)F_9
(x_{\tilde q}, x_b)] + \frac{m^2_\tau}{M^2_W-m^2_\tau}~ [x_cF_4
(x_c,x_\tau)F_8 (x_{\tilde q}, x_b) \nonumber\\ && - F_4 (x_c,x_\tau)F_9
(x_{\tilde q}, x_b) - F_5 (x_c,x_\tau)F_9 (x_{\tilde q}, x_b)] \Big] 
\end{eqnarray}
$M_{\nu}(2,3)$: Diagram 2(c) with $\phi$
\begin{eqnarray}
&& \lambda'_{223}\lambda'_{323}~ g^2 \frac{1}{(16\pi^2)^2} ~V^*_{cb} \Big[
\Big( \frac{x_c - x_s}{x_w - x_\mu} m_{\mu} F_5 (x_c,x_\mu) F_{13} (x_{\tilde
q},x_s) \nonumber\\ &&+ \frac{x_s}{x_w - x_\mu} m_{\mu} [F_7 (x_c,x_\mu)
F_{13} (x_{\tilde q},x_s) - F_6 (x_c,x_\mu) F_{12} (x_{\tilde q},x_s)] \Big)
\nonumber\\ && + \Big( \frac{x_c - x_s}{x_w - x_\tau} m_{\tau} F_5
(x_c,x_\tau) F_{13} (x_{\tilde q},x_s) \nonumber\\ && + \frac{x_s}{x_w -
x_\tau} m_{\tau} [F_7 (x_c,x_\tau) F_{13} (x_{\tilde q},x_s) - F_6
(x_c,x_\tau) F_{12} (x_{\tilde q},x_s)] \Big) \Big]
\end{eqnarray}

\end{document}